\documentclass[twocolumn,twocolappendix,times]{aastex63}

\shorttitle{Quantifying Feedback from Narrow Line Region Outflows}
\shortauthors{Revalski et al.}

\usepackage{float}
\usepackage{color}
\usepackage{comment}
\usepackage{graphics}
\usepackage{epstopdf}
\usepackage{microtype}
\usepackage{subfigure}
\usepackage[normalem]{ulem}
\usepackage{natbib}
\graphicspath{{./}{figures/}}
\usepackage{amsmath}

\newcommand{\lbol}{\ensuremath{L_{bol}}}
\newcommand{\mbh}{\ensuremath{M_{BH}}}
\newcommand{\ledd}{\ensuremath{\lbol/L_{Edd}}}

\definecolor{malachite}{rgb}{0.04, 0.85, 0.32}

\usepackage{soul,xcolor}
\setstcolor{red}

\begin{document}

\title{Quantifying Feedback from Narrow Line Region Outflows in Nearby Active Galaxies. V. The Expanded Sample}

\correspondingauthor{Mitchell Revalski}
\email{mrevalski@stsci.edu}

\author[0000-0002-4917-7873]{Mitchell Revalski}
\affiliation{Space Telescope Science Institute, 3700 San Martin Drive, Baltimore, MD 21218, USA}

\author[0000-0002-6465-3639]{D. Michael Crenshaw}
\affiliation{Department of Physics and Astronomy, Georgia State University, 25 Park Place, Suite 605, Atlanta, GA 30303, USA}

\author[0000-0001-5862-2150]{Garrett E. Polack}
\affiliation{Department of Physics and Astronomy, Georgia State University, 25 Park Place, Suite 605, Atlanta, GA 30303, USA}

\author[0000-0002-9946-4731]{Marc Rafelski}
\affiliation{Space Telescope Science Institute, 3700 San Martin Drive, Baltimore, MD 21218, USA}
\affiliation{Department of Physics and Astronomy, Johns Hopkins University, Baltimore, MD 21218, USA}

\author[0000-0002-6928-9848]{Steven B. Kraemer}
\affiliation{Institute for Astrophysics and Computational Sciences, Department of Physics, The Catholic University of America, Washington, DC 20064, USA}

\author[0000-0002-3365-8875]{Travis C. Fischer}
\affiliation{AURA for ESA, Space Telescope Science Institute, 3700 San Martin Drive, Baltimore, MD 21218, USA}

\author[0000-0001-8658-2723]{Beena Meena}
\affiliation{Department of Physics and Astronomy, Georgia State University, 25 Park Place, Suite 605, Atlanta, GA 30303, USA}

\author[0000-0003-2450-3246]{Henrique R. Schmitt}
\affiliation{Naval Research Laboratory, Washington, DC 20375, USA}

\author[0000-0001-8112-3464]{Anna Trindade Falc\~{a}o}
\affiliation{Harvard-Smithsonian Center for Astrophysics, 60 Garden St., Cambridge, MA 02138, USA}

\author[0000-0001-7238-7062]{Julia Falcone}
\affiliation{Department of Physics and Astronomy, Georgia State University, 25 Park Place, Suite 605, Atlanta, GA 30303, USA}

\author[0009-0005-3001-9989]{Maura Kathleen Shea}
\affiliation{Department of Physics and Astronomy, Georgia State University, 25 Park Place, Suite 605, Atlanta, GA 30303, USA}


\begin{abstract}
We present spatially-resolved measurements of the ionized gas masses and mass outflow rates for six low-redshift ($z \leq$~0.02) active galaxies. In this study, we expand our sample to galaxies with more complex gas kinematics modeled as outflows along a galactic disk that is ionized by the active galactic nucleus (AGN) bicone. We use Hubble Space Telescope (HST) Space Telescope Imaging Spectrograph (STIS) spectroscopy, Wide Field Camera 3 (WFC3) narrow-band imaging, and the photoionization modeling technique that we developed in \cite{Revalski2022} to calculate ionized gas masses using the [O~III]/H$\beta$ ratios at each radius. We combine these with existing kinematic models to derive mass and energy outflow rates, which exhibit substantial radial variations due to changes in the outflow velocities. The full sample of 12 galaxies from this series of studies spans 10$^3$ in bolometric luminosity, and we find that the outflows contain ionized gas masses of $M \approx 10^{4.6} - 10^{7.2}$ $M_{\odot}$, reach maximum mass outflow rates of $\dot M_{out} \approx 0.1 - 13$ $M_{\odot}$ yr$^{-1}$, encompass kinetic energies of $E \approx 10^{52} - 10^{56}$ erg. These energetic properties positively correlate with AGN luminosity. The outflow energetics are less than benchmarks for effective feedback from theoretical models, but the evacuation of gas and injection of energy may still generate long term effects on star-formation in these nearby galaxies. These results highlight the necessity of high spatial resolution imaging and spectroscopy for accurately modeling ionized outflows in active galaxies.
\end{abstract}


\section{Introduction}

\subsection{Feedback from Mass Outflows in Active Galaxies}

Most galaxies host a supermassive black hole (SMBH) that can generate radiation across the electromagnetic spectrum when it is accreting as an active galactic nucleus (AGN). This radiation can drive powerful outflows of ionized, neutral, and molecular gas that may evacuate reservoirs of star-forming gas from the bulge and regulate the rate of SMBH growth \citep{Ciotti2001, Hopkins2005, Kormendy2013, Heckman2014, Fiore2017, Cresci2018, Harrison2018, Storchi-Bergmann2019, Veilleux2020, Florez2021, Laha2021}. Mass outflows are multi-phase and multi-scale in nature, with ionized gas in the narrow-line region (NLR) spanning $\sim$1 -- 1000+ parsecs (pcs) from the central SMBH, and with gas densities of $n_\mathrm{H} \approx 10^1$ -- $10^6$~cm$^{-3}$ \citep{Osterbrock2006, Revalski2022}. These outflows extend from the SMBH's sphere of influence to galactic scales, transporting mass and energy to larger radii within the host galaxy. In this series of papers, we are investigating whether or not NLR outflows have sufficient power to significantly impact their host galaxies by measuring outflow gas masses ($M$) and velocities ($v$) as a function of distance from their SMBHs. These measurements are then used to calculate radial flow rates of gas mass ($\dot M = M v / \delta r$), energy ($\dot E = 1/2 \dot M v^2$), and momentum ($\dot p = \dot M v$) for the NLR outflows in each galaxy.

\begin{deluxetable*}{lcccccccc}
\vspace{-0.5em}
\setlength{\tabcolsep}{0.14in}
\def\arraystretch{0.95}
\tablecaption{Physical Properties of the Active Galaxy Sample}
\tablehead{
\colhead{Catalog} & \colhead{Seyfert} & \colhead{Redshift} & \colhead{Distance} &\colhead{Scale} & \colhead{Inclination} & \colhead{$\log$(\lbol)} & \colhead{$\log$(\mbh)} & \colhead{\ledd \vspace{-0.7em}}\\
\colhead{Name} & \colhead{Type} & \colhead{(unitless)} & \colhead{(Mpc)} &\colhead{(pc/$\arcsec$)} & \colhead{(deg)} & \colhead{(erg s$^{-1}$)} & \colhead{($M_{\odot}$)} & \colhead{(unitless) \vspace{-0.7em}} \\
\colhead{(1)} & \colhead{(2)} & \colhead{(3)} &\colhead{(4)} & \colhead{(5)} & \colhead{(6)} & \colhead{(7)} & \colhead{(8)} & \colhead{(9)}
}
\startdata
IC 3639  & 2   & 0.0109 & 44.9 & 217 & 32.8 & 44.0 & 7.0 & 0.087 \\
NGC 788 & 2   & 0.0136 & 55.8 & 271 & 31.8 & 44.2 & 7.5 & 0.036 \\
NGC 1667 & 2 & 0.0153 & 62.7 & 304 & 45.6 & 43.6 & 7.8 & 0.005 \\
NGC 3393 & 2   & 0.0125 & 51.4 & 249 & 33.9 & 44.8 & 7.7 & 0.107 \\
NGC 7682 & 2 & 0.0171 & 70.4 & 341 & 47.2 & 44.3 & 7.3 & 0.072 \\
UGC 1395 & 1.9    & 0.0172 & 70.6 & 342 & 50.2 & 43.7 & 6.7 & 0.072 \\
\enddata
\tablecomments{Columns are (1) target name, (2) Seyfert classification type, (3) 21~cm redshift from the NASA/IPAC Extragalactic Database, (4) Hubble distance and (5) spatial scale assuming H$_0$ = 73 km s$^{-1}$ Mpc$^{-1}$ \citep{Riess2022}, (6) host galaxy inclination and (7) bolometric luminosity from \cite{Polack2024}, (8) supermassive black hole mass, and (9) the Eddington ratio calculated using $L_{Edd} = 1.26 \times 10^{38} \left(M_{BH}/M_{\odot}\right)$ erg~s$^{-1}$. The black hole masses in column 8 are from \citealp{Fischer2014} and references therein.}
\label{tab:sample}
\end{deluxetable*}

We previously made these measurements for six low-redshift ($z \leq$~0.05) AGN using multi-component photoionization models applied to spatially-resolved spectroscopy covering dozens of emission lines, with the results summarized in \cite{Revalski2021}. This modeling approach eliminated assumptions about the outflow locations, velocities, and densities that can introduce systematic uncertainties of 1~$-$~3 dex \citep{Karouzos2016, Bischetti2017, Perna2017} when calculating a single global value for each energy flow rate based on integrated spectroscopy. However, the above approach requires high signal-to-noise (S/N) spectroscopy covering a wide range in wavelength, and in \cite{Revalski2022} we simplified this technique to reliably calculate ionized gas masses and outflow rates using single component photoionization models with measurements of only the H$\alpha$ $\lambda$6563~\AA, H$\beta$ $\lambda$4861~\AA, and [O III]~$\lambda$5007~\AA~emission lines. While we consider the masses estimated from multi-component models to be the gold standard, single component photoionization models can reproduce those masses to within factors of 2~$-$~4.

\begin{figure*}
\centering
\vspace{-1em}
\includegraphics[width=0.45\textwidth, trim={8.0em 0em 0em 2em}, clip]{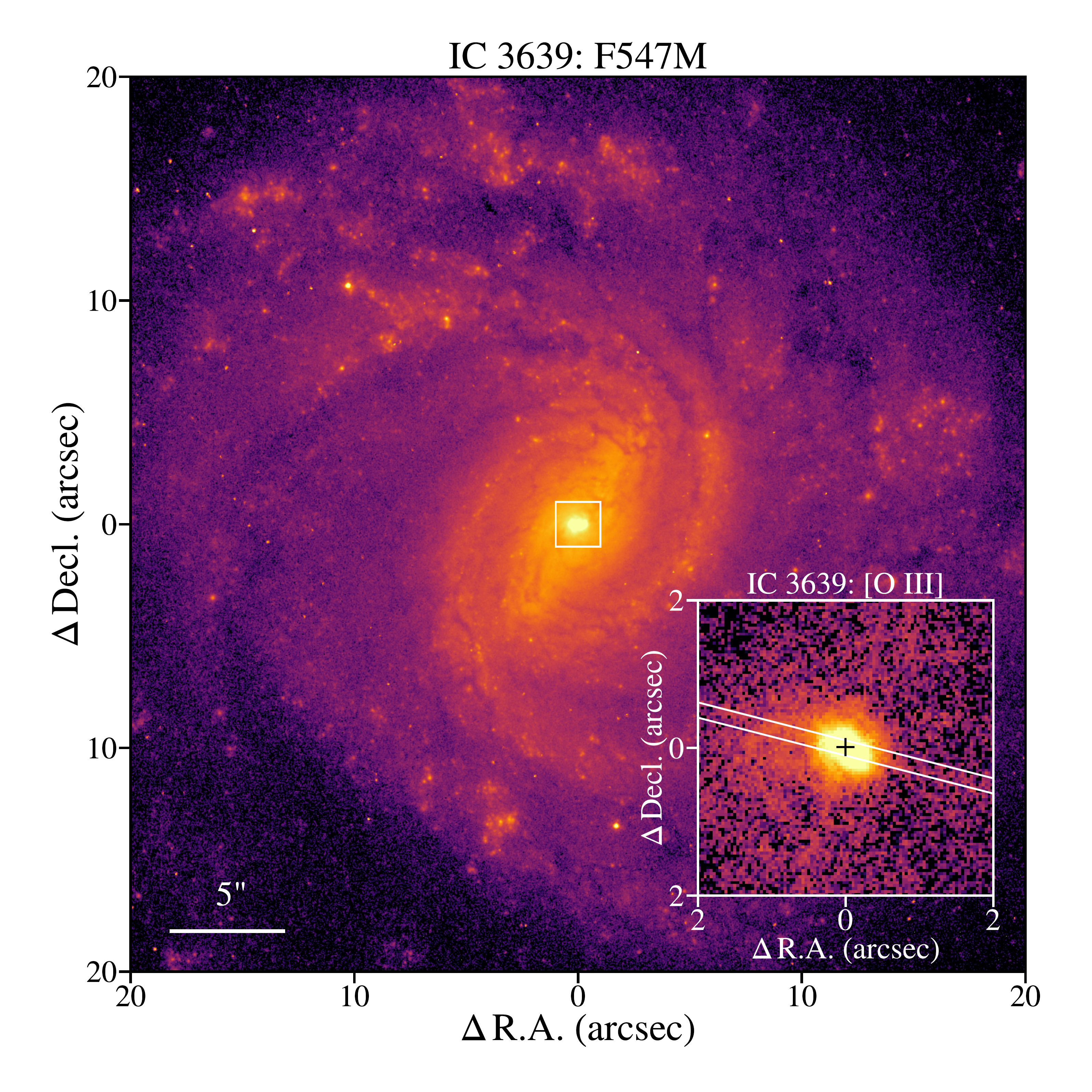}
\includegraphics[width=0.45\textwidth, trim={8.0em 0em 0em 2em}, clip]{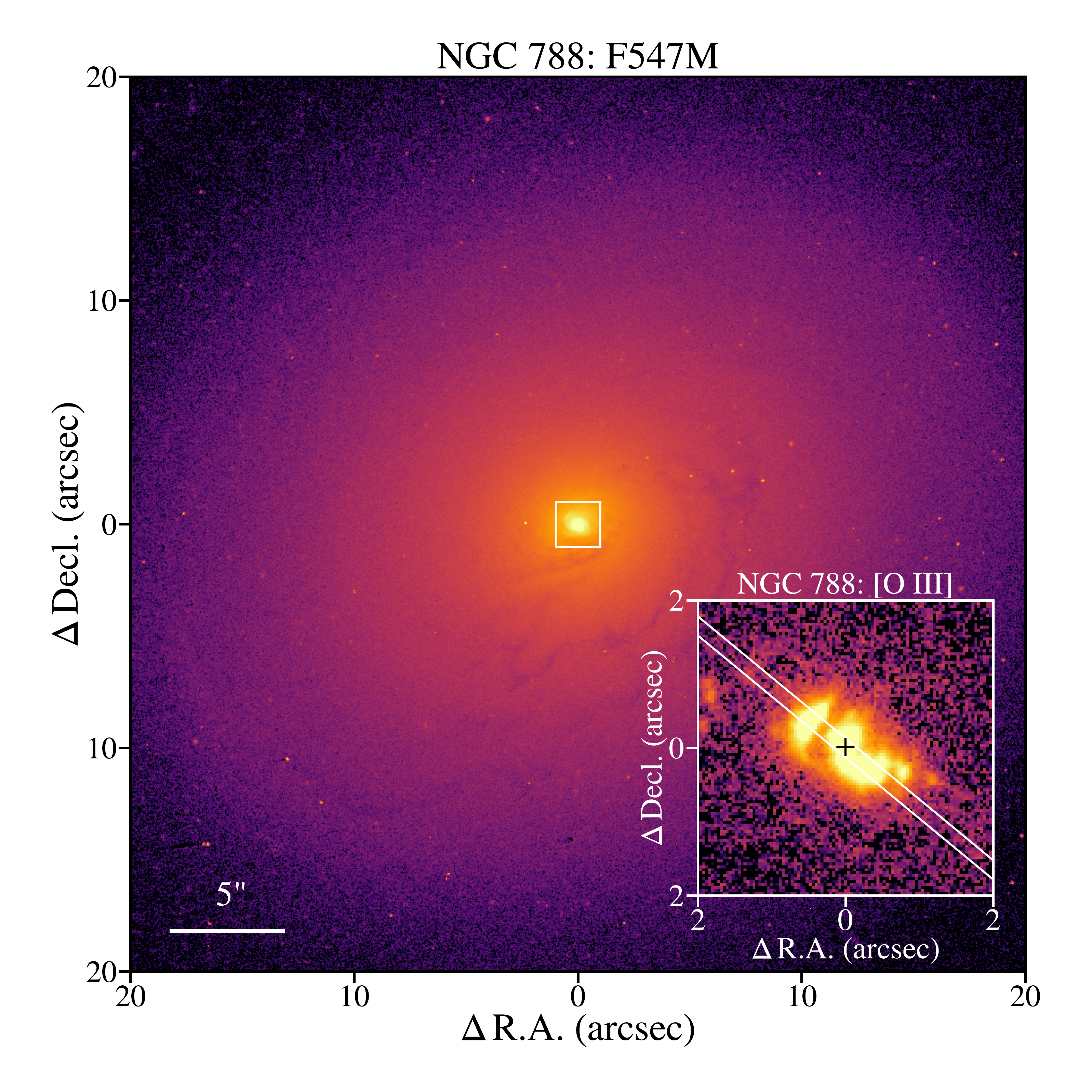}\\\vspace{-1.0em}
\includegraphics[width=0.45\textwidth, trim={8.0em 0em 0em 2em}, clip]{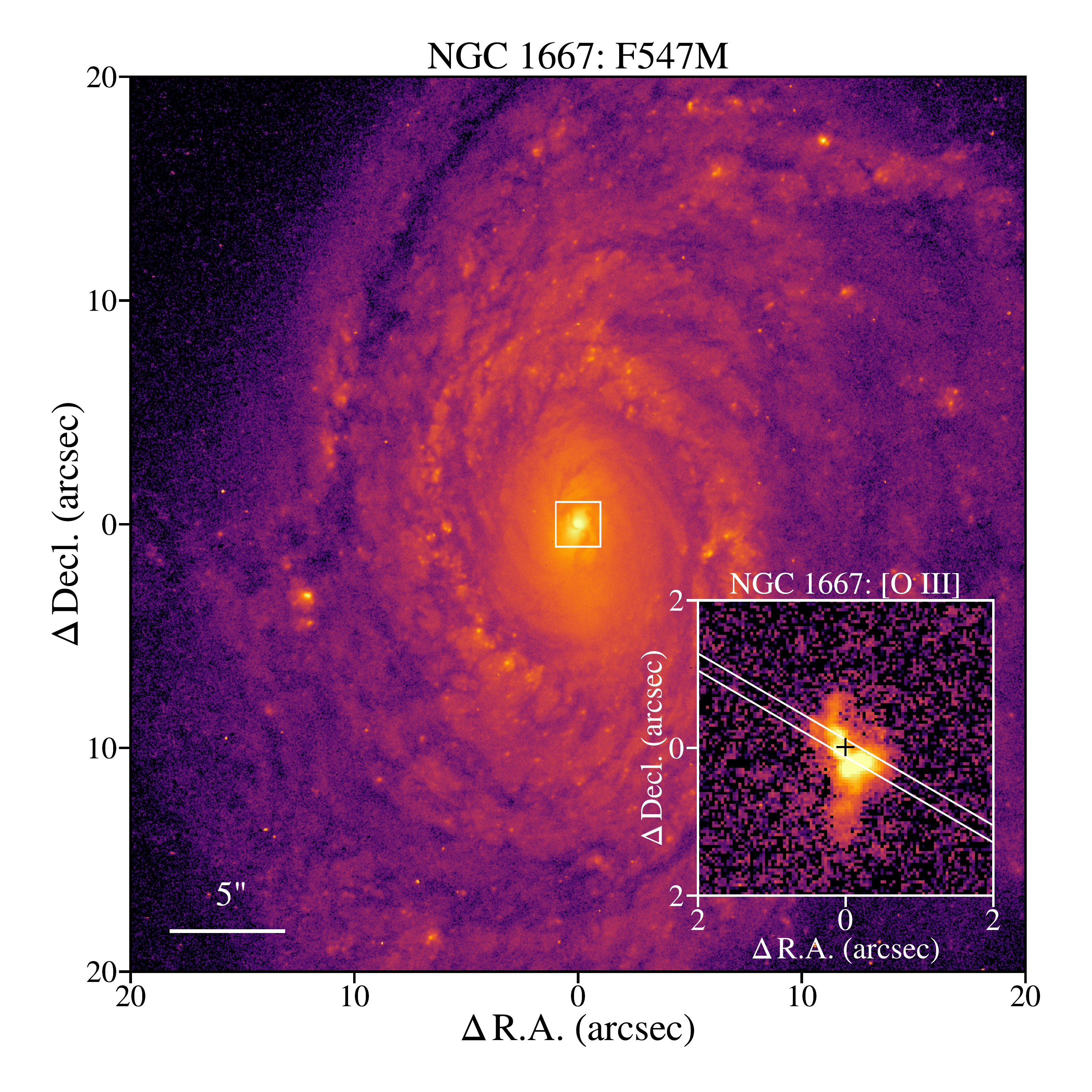}
\includegraphics[width=0.45\textwidth, trim={8.0em 0em 0em 2em}, clip]{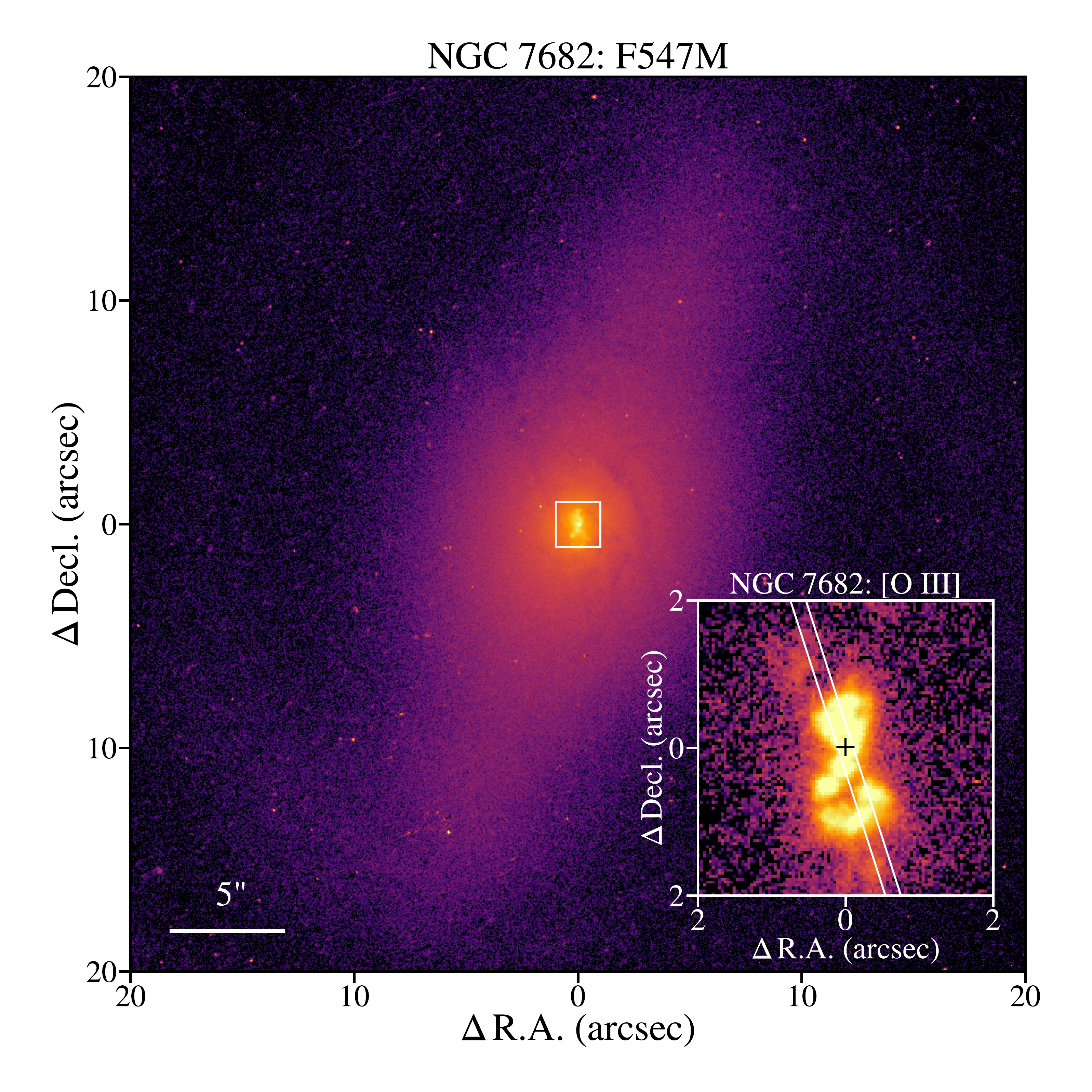}\\\vspace{-1.0em}\hspace{2.35em}
\includegraphics[width=0.45\textwidth, trim={8.0em 0em 0em 2em}, clip]{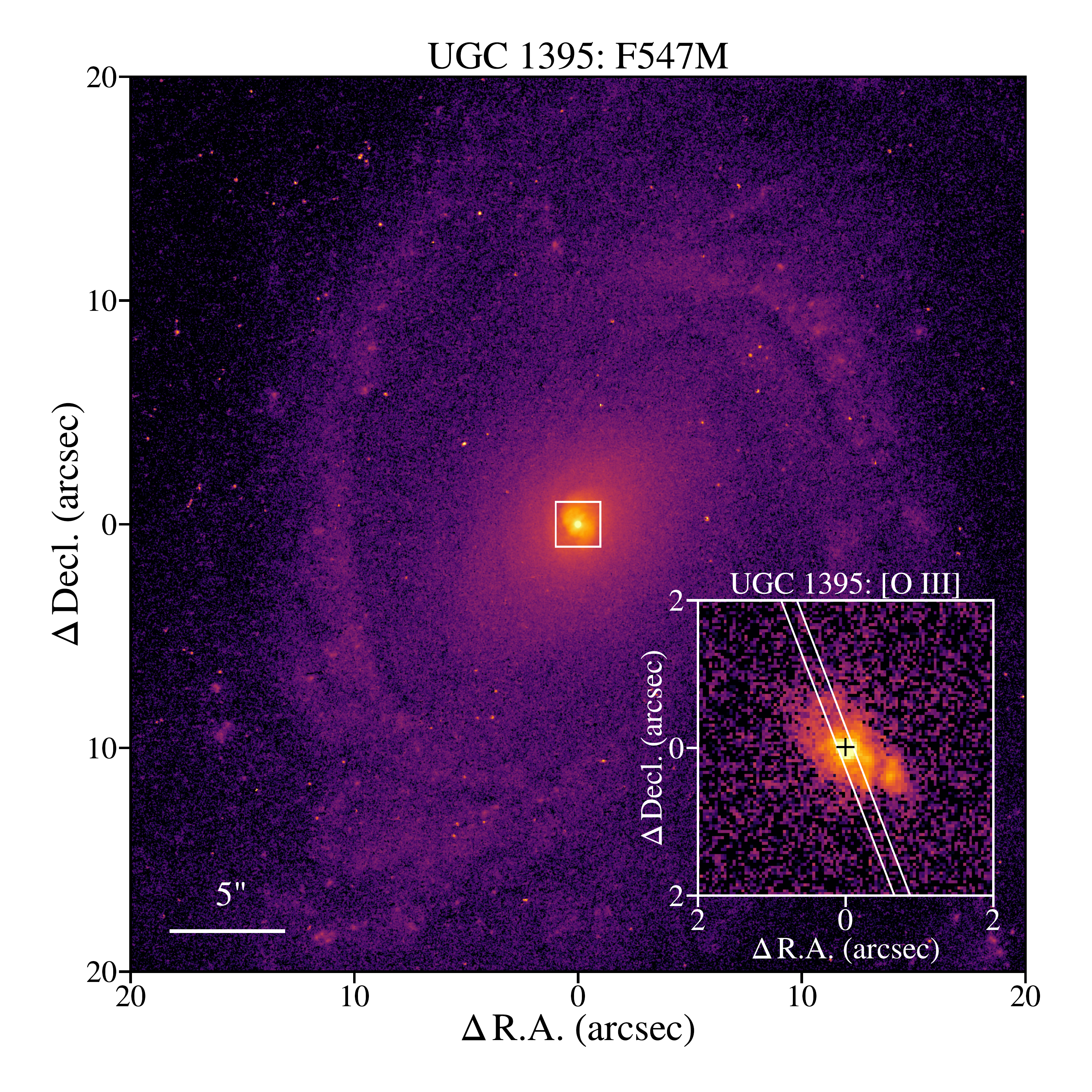}
\begin{minipage}{0.5\textwidth}
\vspace{-27.65em}\hspace{1.3em}
\includegraphics[width=0.8125\textwidth, trim={0em 0em 0em 0em}, clip]{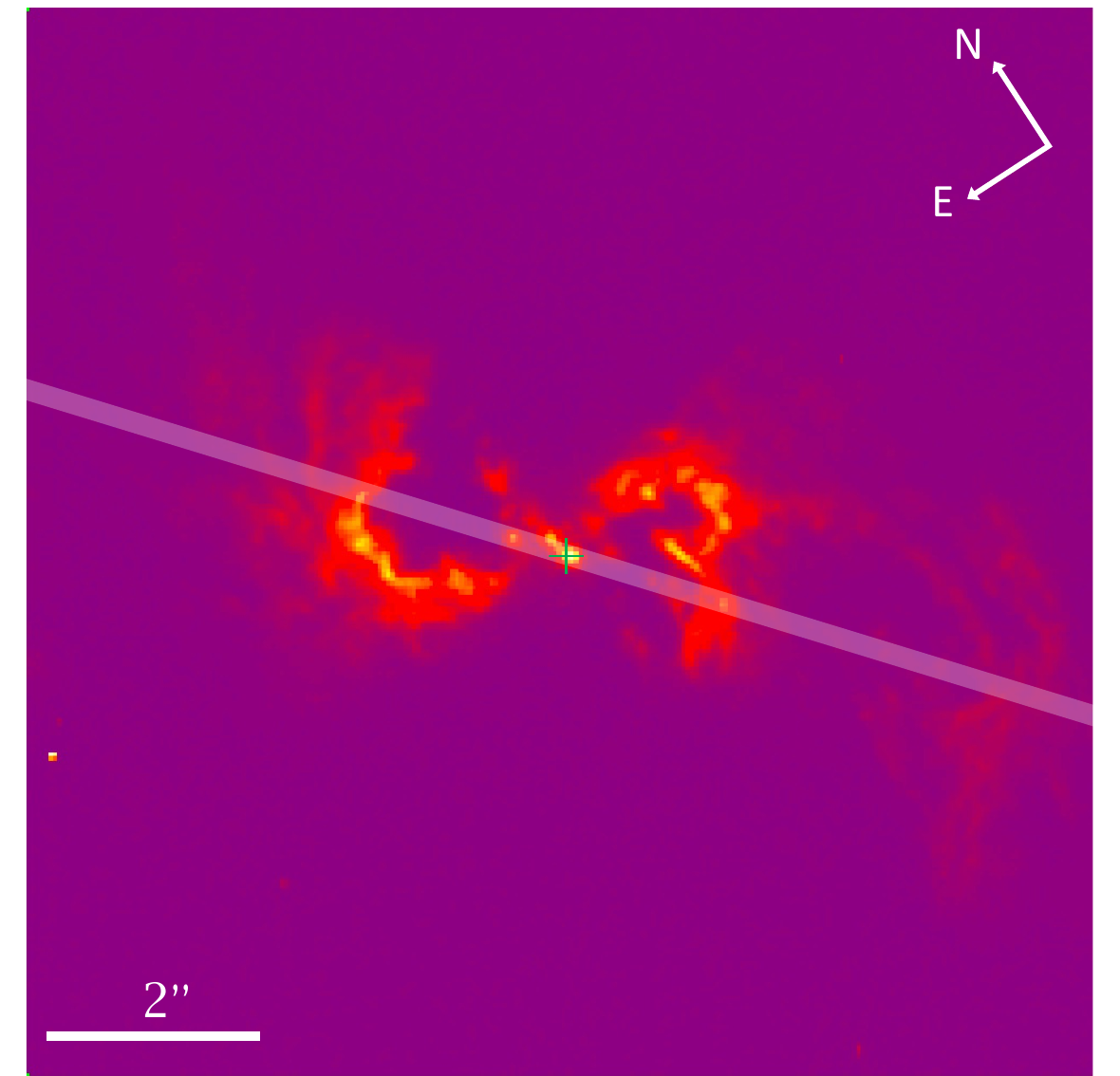}
\end{minipage}\\
\vspace{-2em}
\caption{HST WFC3/UVIS continuum and [O~III] (inset) images for the sample (lower-right is an archival [O~III] image for NGC~3393 from \citealp{Polack2024}). The 40$\arcsec\times$40$\arcsec$ continuum images highlight the galaxy morphology, while the 4$\arcsec\times$4$\arcsec$ [O~III] images zoom in on the ionized gas. The $+$ signs represent the location of the continuum peaks relative to the [O III] emission, and the white lines on the [O~III] insets show the HST STIS long-slit widths (0\farcs2) and position angles. North is up, east is to the left, and the spatial scales in arcseconds pc$^{-1}$ are listed in Table~\ref{tab:sample}.}
\label{fig:sample}
\end{figure*}

\subsection{Sample Selection}

The six galaxies in our earlier study, as well as the six additional galaxies modeled here, were drawn from the original studies by \cite{Fischer2013, Fischer2014}. In those works, the ionized gas kinematics in each galaxy were classified as outflow, ambiguous, complex, or compact. The outflow targets were modeled with a biconical geometry that captured the rise, peak, and decline of the gas kinematics as a function of distance from the nucleus. However, many of the ambiguous or complex sources displayed more chaotic kinematics with clear deviations from rotation, but an uncertain outflow geometry.

In \cite{Polack2024}, we modeled these more chaotic sources with a model consisting of a rotating galactic disk plus outflows driven radially along the disk in its intersection with the AGN ionizing bicone. Thus, with the lack of sufficient constraints on the geometry of the bicone in these sources, we assume that the outflows are confined to the disk, where the gas reservoirs are located \citep{Fischer2017, Gnilka2020}.
Specifically, we re-examined the sources from \cite{Fischer2013, Fischer2014} using the same STIS spectroscopic data, as well as new HST WFC3 narrow-band imaging of the [O III] emission across the full extent of the NLR (HST Program ID \href{https://www.stsci.edu/hst/phase2-public/16246.pdf}{16246}, PI: M. Revalski, \citealp{Revalski2020}). We discovered that many of the targets have STIS long-slit position angles that poorly sample the NLR emission. As a result, we were only able to model six galaxies with satisfactory STIS spectral coverage of their NLR outflows, and we select those six targets for mass outflow modeling in the current study. The physical properties of these galaxies are summarized in Table~\ref{tab:sample}. We show the HST WFC3/UVIS continuum and [O~III] images, first presented in \cite{Polack2024}, in Figure~\ref{fig:sample}. We note that NGC~3227 also has suitable data for this type of study, which due to its multiwavelength complexity is being analyzed separately (Falcone et al., in preparation).

In this study, we model the ionized gas masses and outflow rates for the six galaxies presented in \cite{Polack2024}, which together with the six galaxies modeled in \cite{Revalski2021} represent nearly all of the low-redshift Seyfert galaxies from \cite{Fischer2013, Fischer2014} with the required HST STIS spectroscopy and [O~III] imaging data that are needed for our spatially-resolved mass outflow modeling techniques. Critically, these six AGN double the sample of galaxies with outflow energetics determined from spatially-resolved photoionization models, which involve fewer assumptions and alleviate biases as compared to commonly used global techniques that average over the spatially extended outflow properties \citep{Revalski2022}.

\newpage
\section{Data \& Methods}
\label{sec:methods}

\subsection{HST Observations}

The HST WFC3 image reduction, STIS spectral fitting, and kinematic modeling for the sample were presented in \cite{Polack2024}, and we highlight the key details here for completeness. Specifically, we observed the sample with HST WFC3/UVIS in Cycle 28, obtaining continuum images with the F547M filter, and narrow-band images with the FQ508N or F502N filter to sample the [O~III] $\lambda$5007~\AA~emission line. We utilized a single orbit for each galaxy, with details of the observational setup and mosaic drizzling provided in \cite{Polack2024}. We generated the continuum-subtracted [O~III] images by subtracting off the continuum measured in the F547M filter, scaled down to the width of the narrow band filters, from their corresponding F502N or FQ508N mosaics. The WFC3 observations can be accessed using the DOI: \dataset[10.17909/np2s-gd34]{https://doi.org/10.17909/np2s-gd34}. We also provide our custom calibrated mosaics for use by the community as High Level Science Products available from the Mikulski Archive for Space Telescopes (MAST) using the DOI: \dataset[10.17909/7ccz-mc93]{https://doi.org/10.17909/7ccz-mc93}, and directly online at \textbf{\url{https://archive.stsci.edu/hlsp/nlr-agn/}}. The HST WFC3 continuum and [O~III] images are shown in Figure~\ref{fig:sample}.

We utilized archival HST STIS spectroscopy for measuring the gas kinematics and emission line ratios, primarily with the G750M and G430L gratings covering the H$\alpha$ and [O~III] portions of the optical spectrum, respectively. The observation and program ID's, exposure times, and slit position angles are provided in Table~2 of \cite{Polack2024}. We fit the emission lines with a Bayesian routine known as BEAT \citep{Fischer2017}, which determines the most probable number of Gaussians needed to model the line profile, corresponding to unique kinematic components of gas. An example of the resulting fits are shown in Figure~3 of \cite{Revalski2021}.

\subsection{Kinematic Models}

Using this framework, \cite{Polack2024} fit the emission line profiles along each STIS slit to measure the observed emission line centroids, full widths at half maximum (FWHM), and fluxes. They then adopted a kinematic model where gas is considered in rotation if the line velocity and width are $<$ 250 km s$^{-1}$, disturbed if the centroid is $<$ 250 km s$^{-1}$ but the FWHM is $>$ 250 km s$^{-1}$, and outflow if both the line centroid and the FWHM are $>$ 250 km s$^{-1}$. They then deprojected the velocities for the outflowing components, assuming that they are being radially driven outwards along the galaxy disk as discussed above. The deprojected kinematics, distances, and outflow extents are shown in Figure~10 of \cite{Polack2024}. The deprojected outflow velocities as a function of distance from the SMBH are required for calculating the ionized gas mass outflow rates in each AGN.
\\
\subsection{Emission Line Diagnostics}

\begin{figure*}
\centering
\includegraphics[width=0.49\textwidth, trim={0em 0em 0em 0em}, clip]{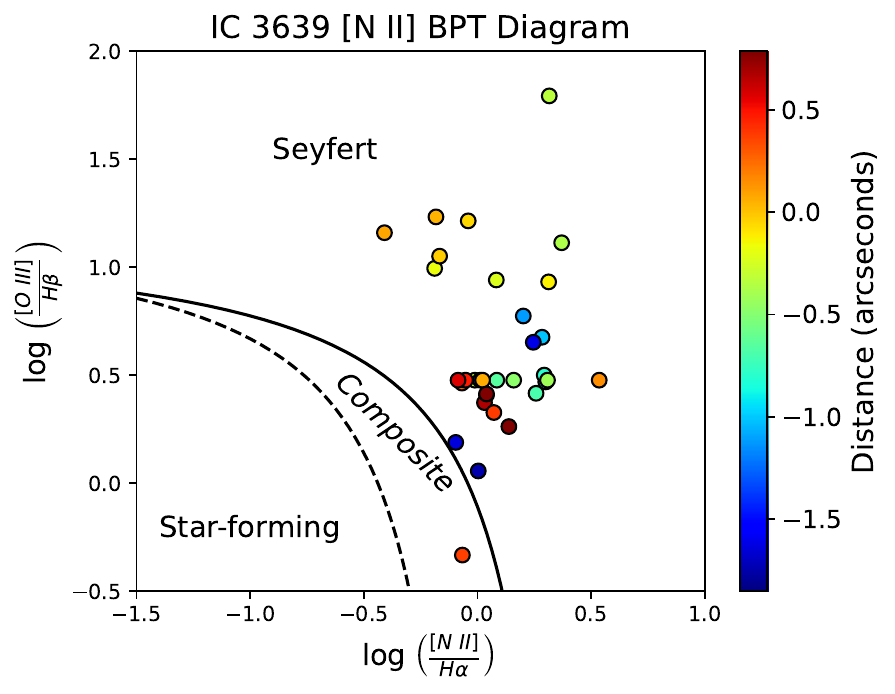}
\includegraphics[width=0.49\textwidth, trim={0em 0em 0em 0em}, clip]{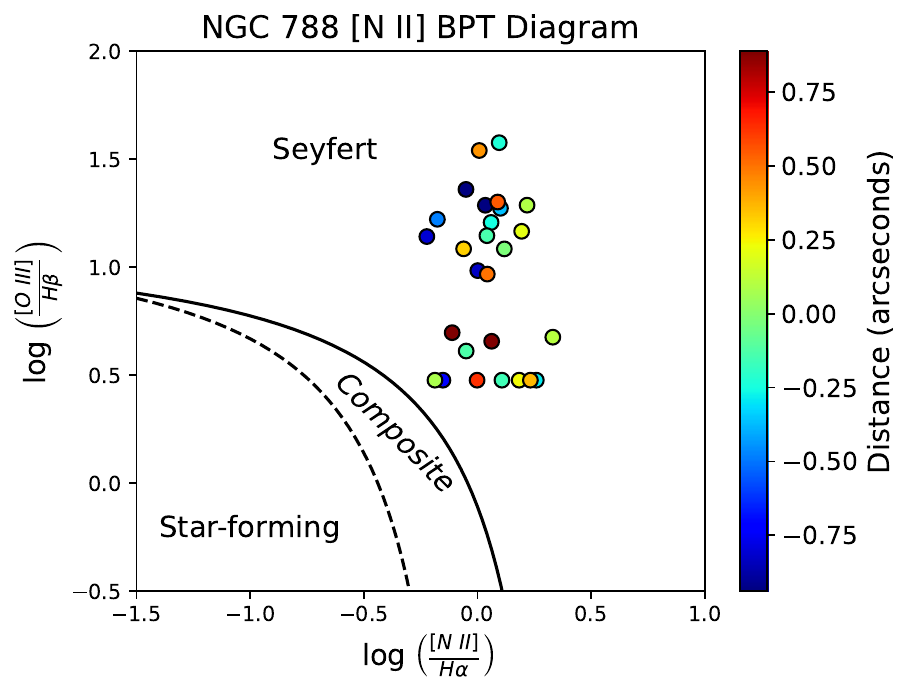}
\includegraphics[width=0.49\textwidth, trim={0em 0em 0em 0em}, clip]{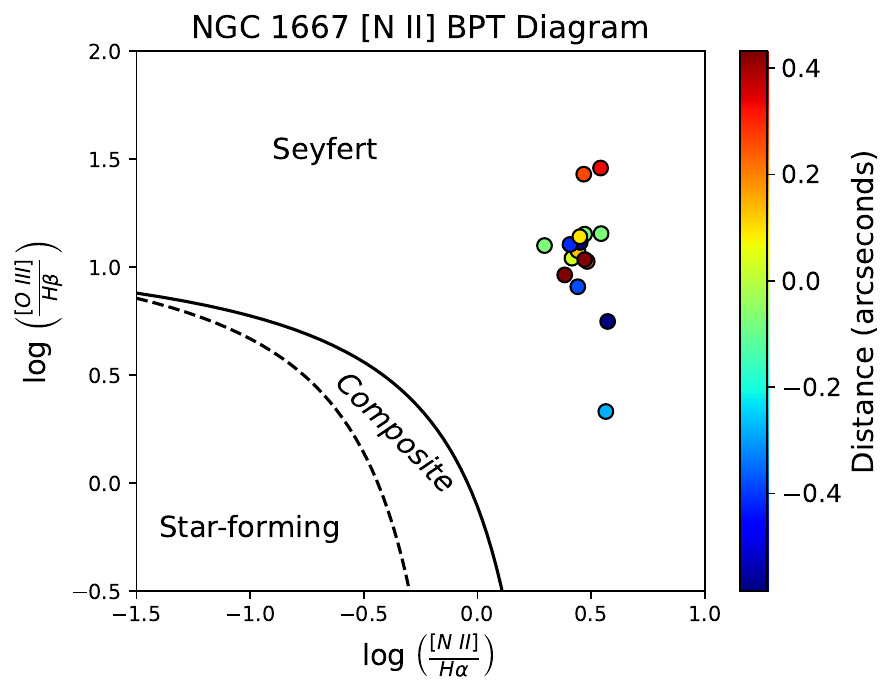}
\includegraphics[width=0.49\textwidth, trim={0em 0em 0em 0em}, clip]{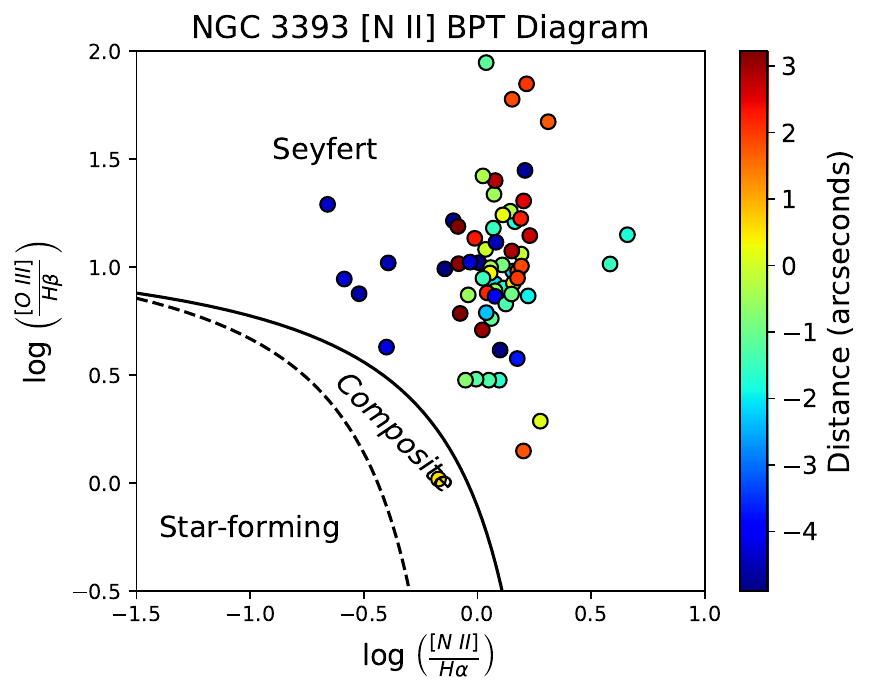}
\includegraphics[width=0.49\textwidth, trim={0em 0em 0em 0em}, clip]{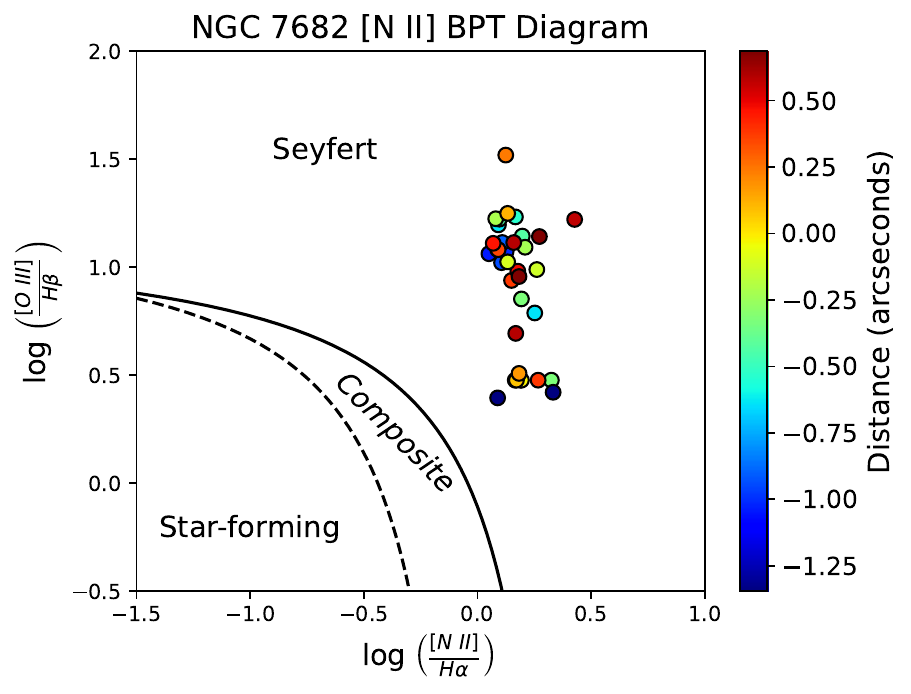}
\includegraphics[width=0.49\textwidth, trim={0em 0em 0em 0em}, clip]{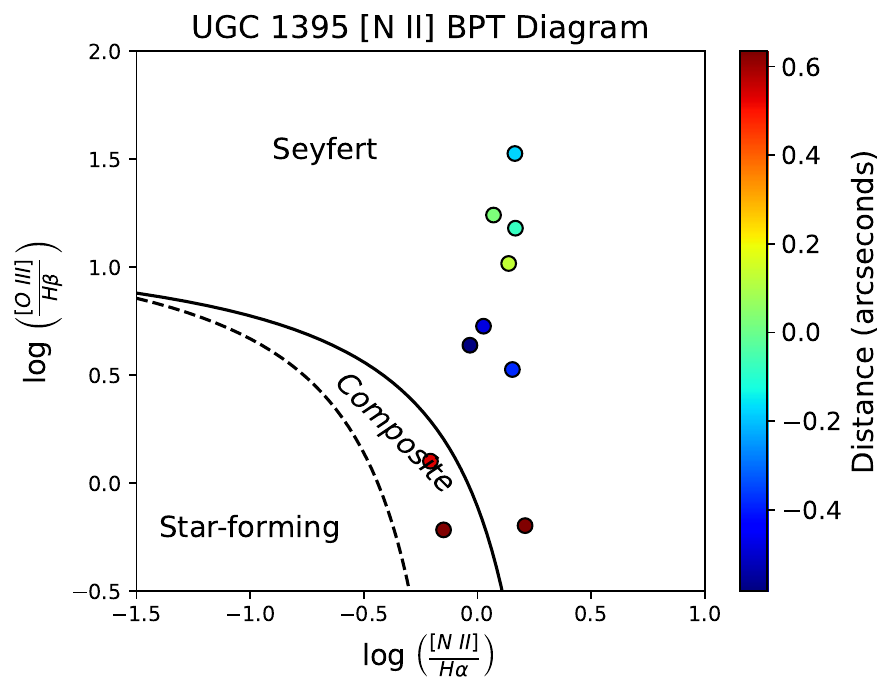}
\caption{The [N~II] BPT diagrams for sources in the sample. This diagnostic differentiates the source of ionization using ratios of lines with different ionization potentials so the ratios vary substantially based on the SED of the ionizing source \citep{Baldwin1981, Veilleux1987}. The separation lines for distinguishing each ionization mechanisms are from \cite{Kewley2001, Kewley2006} and \cite{Kauffmann2003}. The points in each panel are color-coded by their distance from the nucleus, with positive distances corresponding to locations higher in the STIS slits.}
\label{fig:bpt}
\end{figure*}

We utilized our line flux measurements to construct Baldwin-Phillips-Terlevich (BPT) diagnostic diagrams at each spatial location that differentiate sources of ionization using ratios of lines with different ionization potentials such that their ratios vary significantly based on the spectral energy distribution (SED) of the ionizing source \citep{Baldwin1981, Veilleux1987}. In Figure~\ref{fig:bpt}, we present the spatially-resolved [N~II] BPT diagram for each AGN. At all spatial locations the results are consistent with AGN ionization, indicating the AGN's influence extends to the full radial extent of each outflow. The few exceptions are two points in UGC~1395, and one in IC~3639 and NGC~3393 that are consistent with a mix of AGN and star-forming ionization at the extreme radial extents of the measured outflows. Considering these factors, we model all locations with an AGN SED.

\subsection{Photoionization Models}

In \cite{Revalski2021} we derived the ionized gas masses and outflow rates using photoionization models with multiple density components. While these models accurately constrained the gas masses by accounting for a wide range in ionization, this technique required high S/N spectroscopy covering dozens of emission lines over a wide range in wavelength. In \cite{Revalski2022} we used these results as a benchmark, and compared the accuracies of different techniques for estimating gas densities and calculating ionized gas masses in the NLRs of nearby Seyfert galaxies. While commonly used techniques such as assuming a constant gas density at all locations failed to reproduce the multi-component results of \cite{Revalski2021}, we found that single component photoionization models with densities that vary with distance from the SMBH can reproduce the ionized gas masses estimated by the multi-component models to within factors of $\sim$2$-$4. In these cases, the gas density at each radius is constrained by its distance and ionization parameter, where the ionization parameter at each radius is obtained from the [O~III]/H$\beta$ ratio. We adopt this single component modeling approach in the current study, and detail below how the gas densities are constrained using the [O~III]/H$\beta$ ratios.

\begin{figure*}
\centering
\includegraphics[width=0.495\textwidth, trim={2.2em 0em 2em 0em}, clip]{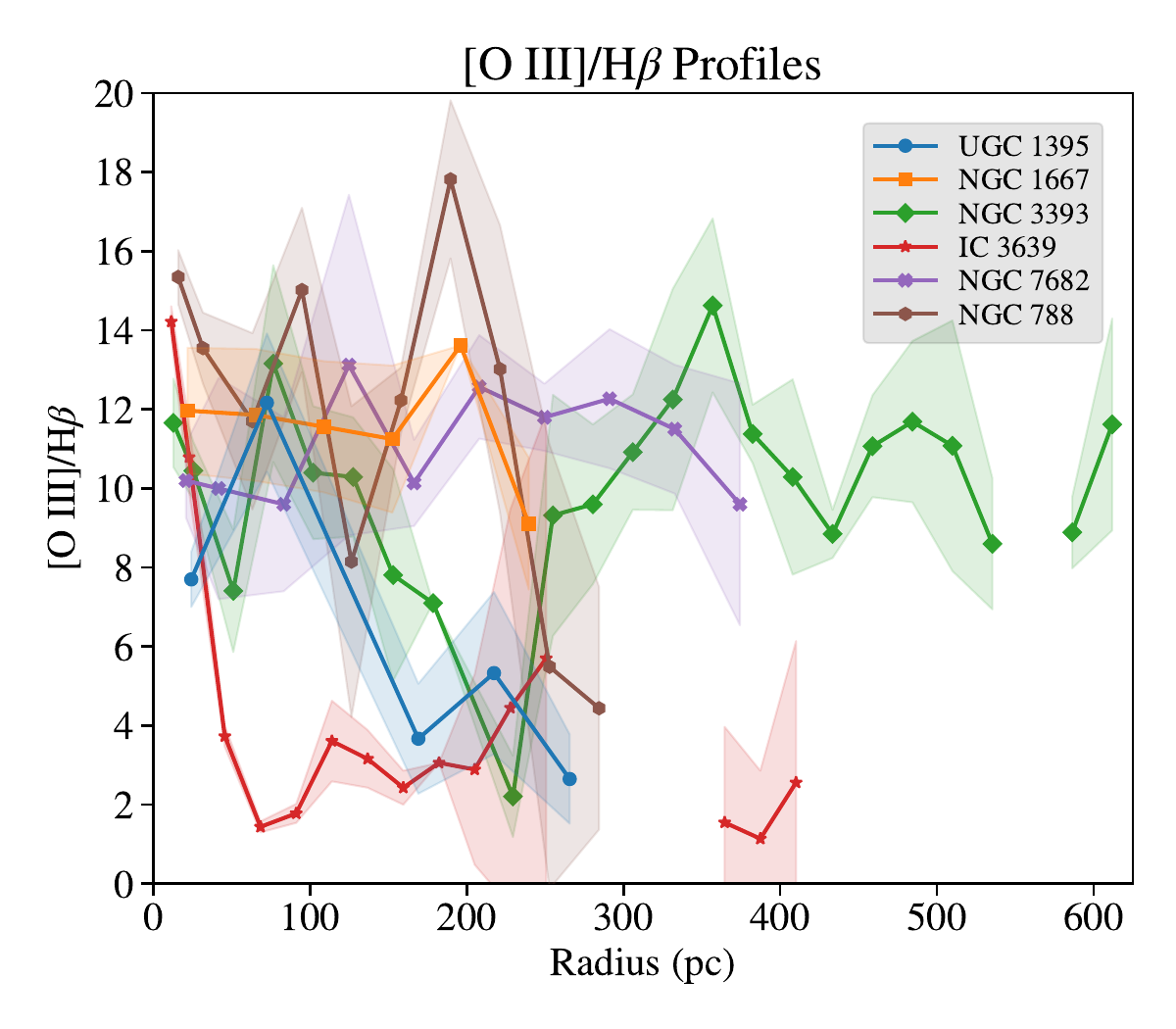}\vspace{-0em}
\includegraphics[width=0.495\textwidth, trim={2.2em 0em 2em 0em}, clip]{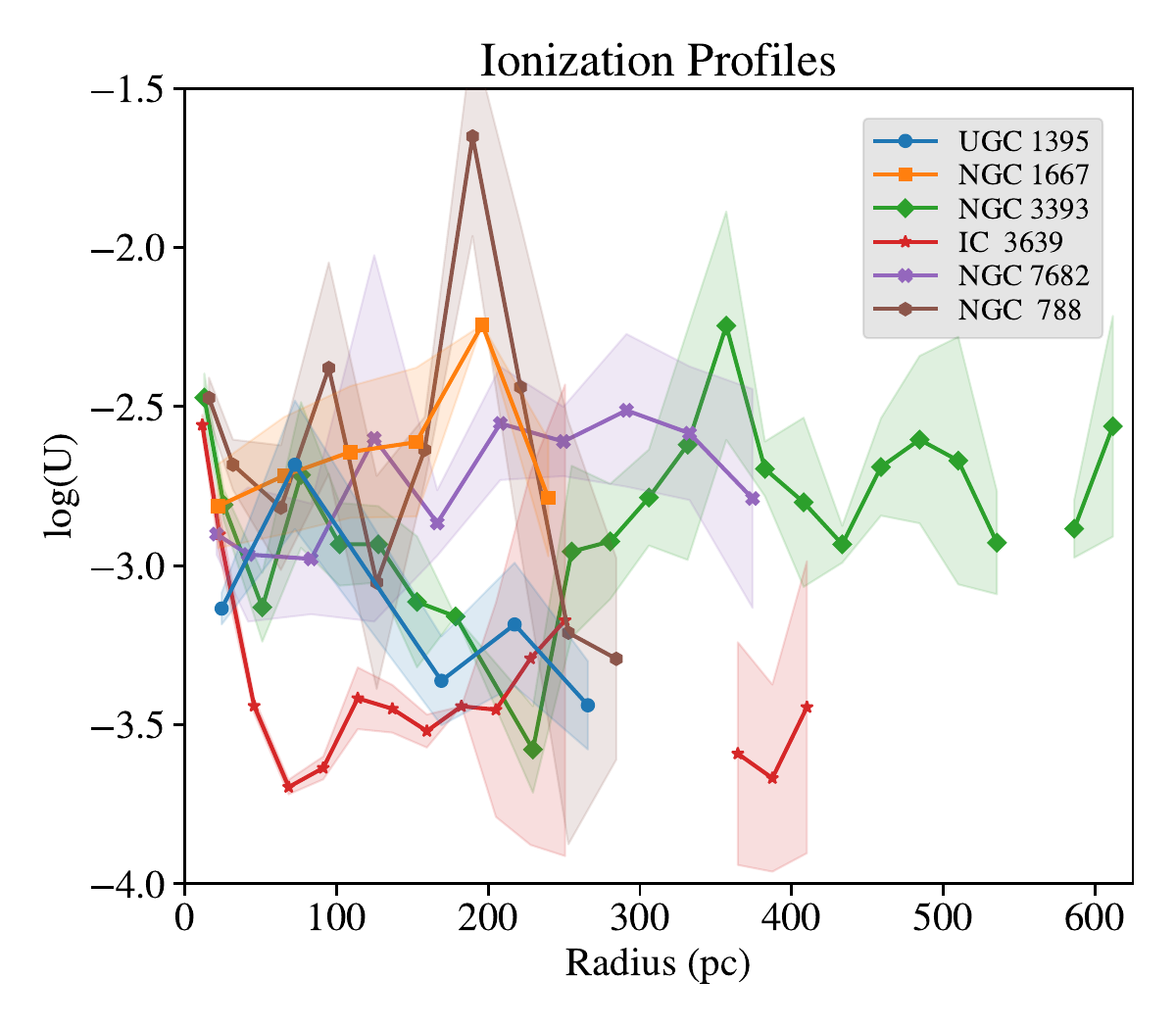}\vspace{-1em}
\includegraphics[width=0.495\textwidth, trim={2.2em 0em 2em 0em}, clip]{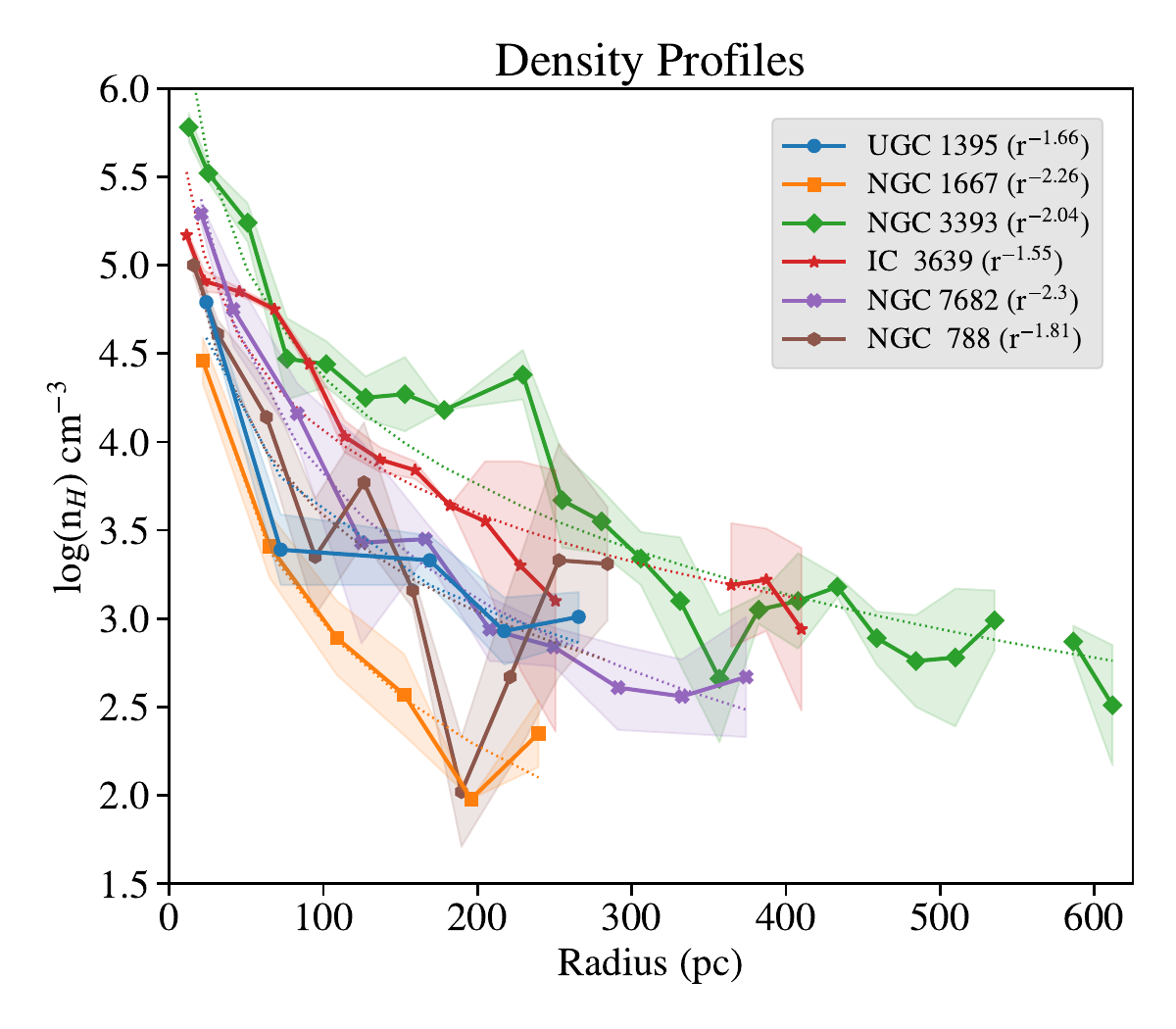}
\includegraphics[width=0.495\textwidth, trim={2.2em 0em 2em 0em}, clip]{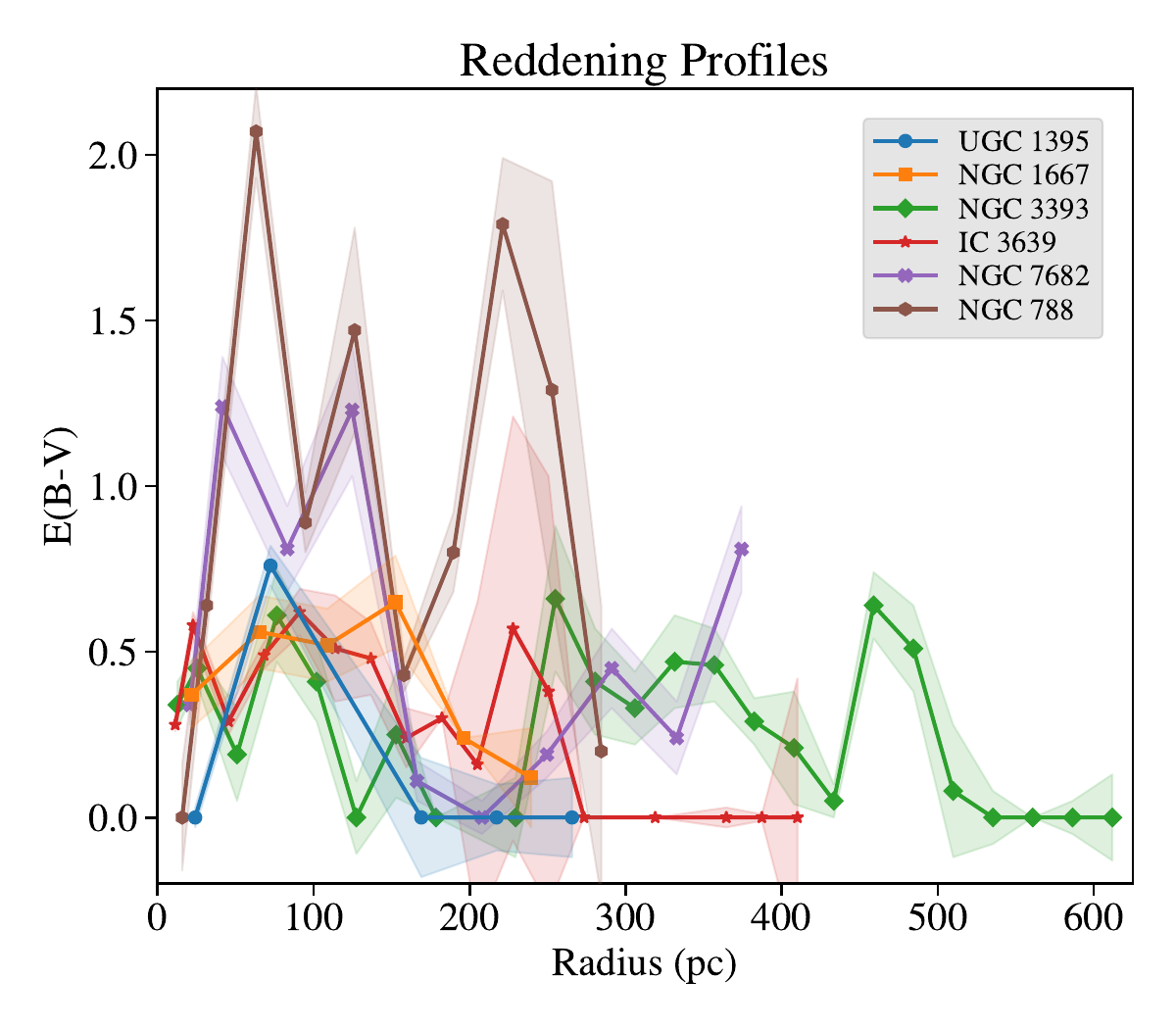}\vspace{-1em}
\caption{The radial [O~III]/H$\beta$ ratios (upper left), ionization parameters ($U$, upper right), and gas densities ($n_\mathrm{H}$, lower left) as functions of distance from the nucleus for each galaxy. The [O~III]/H$\beta$ ratios were converted to values of log($U$) using the Cloudy model grids described \S\ref{sec:methods} (e.g. Figure~5 of \citealp{Revalski2022}). The gas densities were then calculated using these log($U$) values and Equation~\ref{eqn:uion}. Finally, we show values of the color-excess (E(B-V), lower right) in magnitudes as derived from the observed H$\alpha$/H$\beta$ ratios. The shaded regions in the upper left panel denote the errors on the reddening-corrected [O~III]/H$\beta$ ratios, which are propagated from uncertainties in the H$\beta$ and [O~III] fluxes, and E(B-V). The [O~III]/H$\beta$ ratio is largely insensitive to reddening, indicating that the radial variations primarily trace changes in the ionization state and thus density of the gas. In the lower left panel, we show the best fit power law for each density profile, with the power law indices listed in the legend. The small gaps in the radial profiles for IC~3639 and NGC~3393 are generally due to a lack of H$\beta$ detections at each of those radial locations.}
\label{fig:ratios}
\end{figure*}

We generate our models with v23.01 of the Cloudy \citep{Chatzikos2023} photoionization modeling code. Creating self-consistent photoionization models requires knowing the energy distribution of ionizing photons that are intercepting a gas cloud with specified geometry and chemical composition. The AGN luminosity and spectral energy distribution (SED) determine the number of ionizing photons per second, or Q(H)$_{ion}$. We use the luminosities provided in Table~\ref{tab:sample}, and adopt the same broken power-law SED that we used in \cite{Revalski2021}. Namely, $L_{\nu} \propto \nu^{\alpha}$ with slopes of $\alpha$ = --0.5 from 1 eV to 13.6 eV, $\alpha$ = --1.4 from 13.6 eV to 0.5 keV, $\alpha$ = --1 from 0.5 keV to 10 keV, and $\alpha$ = --0.5 from 10 keV to 100 keV, with cutoffs below 1 eV and above 100 keV. Similarly, we assume the gas has a dust level equal to 50\% of that in the Milky Way with corresponding depletions of heavy elements into dust grains \citep{Seab1983, Snow1996, Collins2009}. We use a metallicity of $\sim$1.3~$Z_\odot$ that is typical across many NLRs the abundances are: He~=~--0.96, C~=~--3.63, N~=~--3.94, O~=~--3.32, Ne~=~--3.96, Na~=~--5.65, Mg~=~--4.57, Al~=~--5.70, Si~=~--4.66, P~=~--6.48, S~=~--4.77, Ar~=~--5.49, Ca~=~--5.81, Fe~=~--4.67, Ni~=~--5.93.

\subsection{Ionized Gas Masses}

In this framework, the gas mass at each radial distance is calculated using an optically-thick photoionization model with a single density at each radius utilizing the equation:

\begin{equation}
M_{ion} = N_\mathrm{H} \mu m_p \left(\frac{L_{\mathrm{H}\beta}}{F_{\mathrm{H}\beta_m}}\right),
\label{eqn:mion}
\end{equation}
\noindent
where $N_\mathrm{H}$ is the hydrogen column density (cm$^{-2}$) predicted by the Cloudy model, $\mu =$~1.4 is the mean mass per proton, $m_p$ is the proton mass, $F_{\mathrm{H}\beta_m}$ is the Cloudy model flux for H$\beta$, and $L_{\mathrm{H}\beta}$ is the H$\beta$ luminosity (erg s$^{-1}$) derived from the extinction-corrected flux and galaxy distance. Physically, the ratio of the luminosity and model flux is the area of the clouds, which is multiplied by the column density (cm$^{-2}$) to yield the total number of particles at each location modeled.

The gas density (n$_\mathrm{H}$) is an input parameter for the photoionization models, which yield $N_\mathrm{H}$ and $F_{\mathrm{H}\beta_m}$. The gas density is a most critical model parameter, as it determines the H$\beta$ flux from the model that is directly used in calculating the gas masses (Equation~\ref{eqn:mion}). With spatially-resolved observations the distance of the gas from the AGN can be measured, and so the hydrogen number density ($n_\mathrm{H}$) is fixed for a given ionization parameter ($U$), which is the ratio of the number of ionizing photons to hydrogen atoms at the face of the gas cloud. By inverting the ionization parameter equation \citep{Osterbrock2006, Baron2019a, Baron2019b} the density is:
\begin{equation}
n_\mathrm{H} = \left(\frac{Q(H)_{ion}}{4 \pi r^2 c~U}\right)
\label{eqn:uion}
\end{equation}
\noindent
where $Q(H)_{ion}$ is the number of ionizing photons s$^{-1}$ coming from the AGN, $r$ is the distance of the gas from the AGN, and $c$ is the speed of light. In Figure~5 of \cite{Revalski2022} we presented a grid of Cloudy photoionization models \citep{Ferland2013, Chatzikos2023} showing the strong dependence of the [O~III]/H$\beta$ ratio on the ionization parameter, with minor secondary dependencies on metallicity and gas density over the range of densities typically observed in the NLR. Thus, we use the extinction-corrected [O~III]/H$\beta$ ratio at each radius to determine $U$, and calculate the gas density using Equation~\ref{eqn:uion}. Finally, the standard expressions for converting the observed H$\beta$ flux to an extinction-corrected luminosity are given by Equations~4 and 5 in \cite{Revalski2022}.

The flux-weighted radially averaged [O~III]/H$\beta$ ratios for each galaxy, as well as the corresponding log($U$) and densities, are shown in Figure~\ref{fig:ratios}. Interestingly, most of these galaxies show large radial variations in [O~III]/H$\beta$, unlike our sample in \cite{Revalski2021}. Cases where the [O~III]/H$\beta$ ratio does not change with radius imply a constant ionization parameter, which leads to the gas density decreasing $\propto$~1/r$^{2}$ (i.e. all quantities in Equation~\ref{eqn:uion} are constant except distance). We calculated the best fit power law for the density profile for each source, and report those values in the legend of Figure~\ref{fig:ratios}. Interestingly, the average power law index of 1.94 is exceptionally close to this $\propto$~1/r$^{2}$ law, with relatively small deviations of $\sim$0.3~dex between targets, and larger deviations at specific locations. The largest deviation is seen for IC~3639, where the three points are large radii, with very low [O~III]/H$\beta$ ratios and corresponding high densities, shallow the slope.

We improve on our study in \cite{Revalski2021} by calculating the reddening at each location along the STIS slits from the observed H$\alpha$/H$\beta$ ratios, rather than using an average value across the radial spatial positions. While the [O~III]/H$\beta$ ratio is largely insensitive to reddening due to the small wavelength separation of these lines, we utilize the large wavelength span of our spectra covering H$\alpha$ and H$\beta$ to correct for reddening in order to derive the most accurate line luminosities that directly impact the mass calculations.

We converged on the best-fit models by running a small grid over log(U) and $n_\mathrm{H}$ at each radius, for each galaxy, constrained by Equation~\ref{eqn:uion}. We then interpolate between grid points to find the best-fit log(U) for each [O~III]/H$\beta$ ratio, and then run a final set of models with that log(U) and corresponding density from Equation~\ref{eqn:uion} at each spatial location. This iterative process yields agreement between the data and model [O~III]/H$\beta$ ratios to within 2.5\% at all spatial positions. The small gaps in the [O~III]/H$\beta$ profiles shown in Figure~\ref{fig:ratios} are typically due to a lack of H$\beta$ detections at those locations.

\section{Results}
\label{sec:results}

\begin{figure*}
\centering
\vspace{-1em}
\includegraphics[width=0.7\textwidth, trim={2.5em 0em 2.0em 0em}, clip]{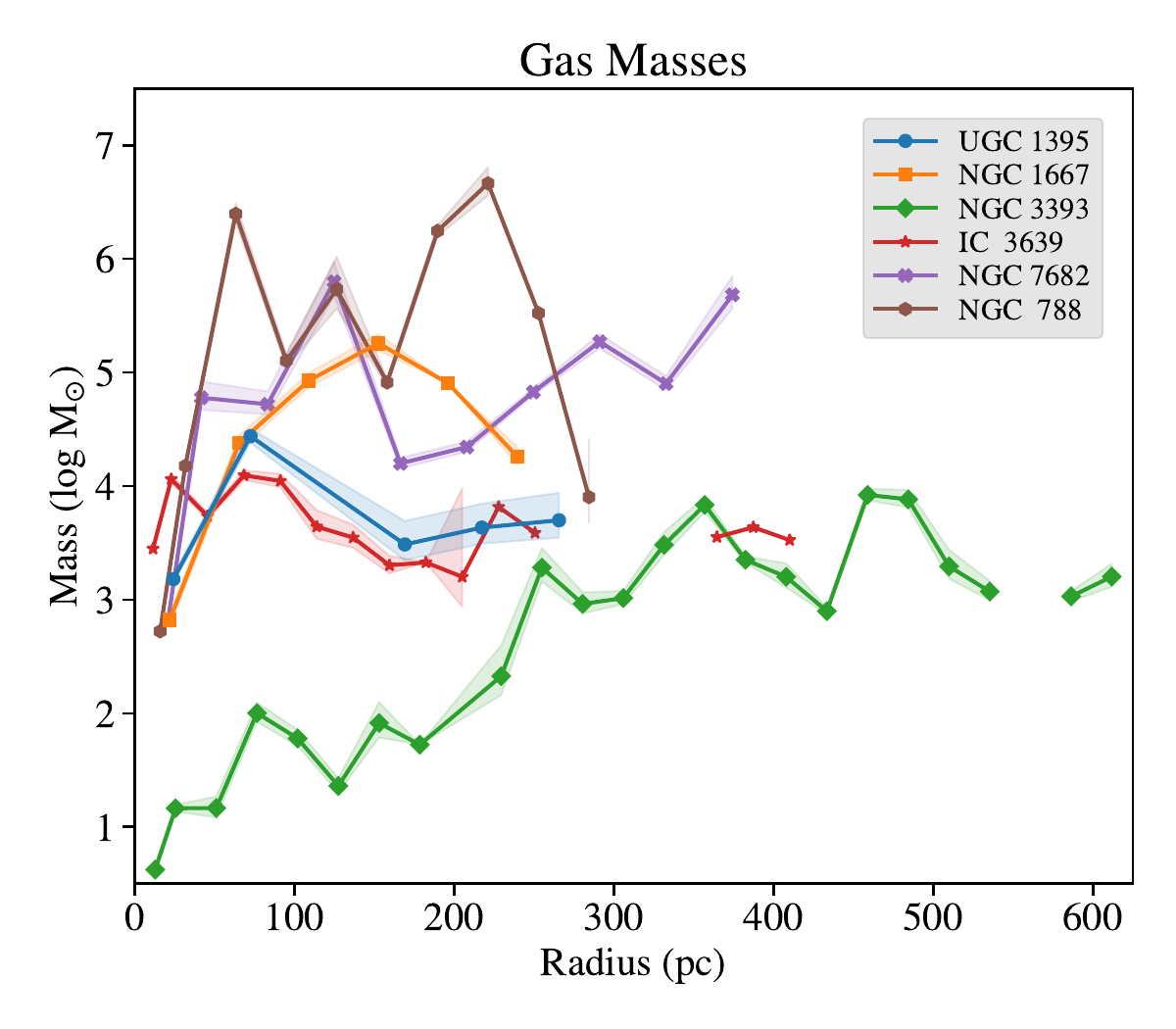}\\\vspace{-1.5em}
\includegraphics[width=0.7\textwidth, trim={2.5em 0em 2.0em 0em}, clip]{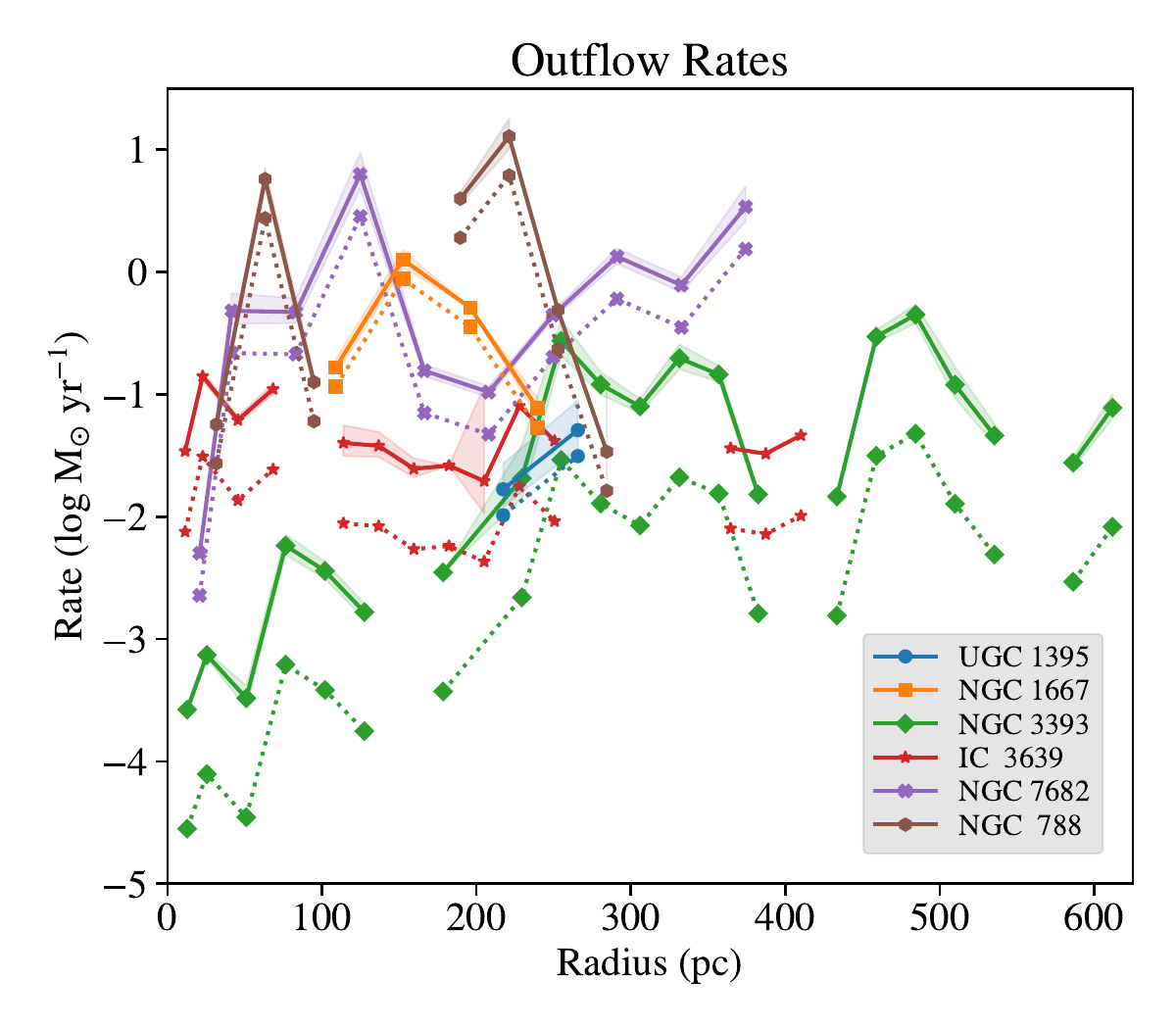}\vspace{-1em}
\caption{The radial gas masses (upper) and mass outflow rates (lower) on logarithmic scales. The outflow rates are calculated using the deprojected velocities as shown by the solid lines, while lower limits for each galaxy calculated using the observed velocities are shown with dotted lines. The shaded regions denote the errors, which are propagated from uncertainties in the H$\beta$ and [O~III] fluxes, and E(B-V) values. The gaps in the outflow rate profiles that are not seen in the mass profiles are because the gas is not outflowing ($v<$~250 km s$^{-1}$) at those locations.}
\label{fig:results}
\end{figure*}

We present radial profiles for the ionized gas masses (per 0\farcs1 bin) and mass outflow rates ($\dot M = M v / \delta r$) in Figure~\ref{fig:results}. We show the outflow rates calculated using the deprojected velocities as solid lines and our observed kinematics with dotted lines. These represent the absolute upper and lower limits, respectively, as they assume that all or none of the velocity vectors are viewed with projection effects. As we found for the six AGN in \cite{Revalski2021}, the gas masses generally increase with distance from the SMBH at small radii as the AGN radiation intercepts more of the host galaxy gas reservoir, but then flatten or even turn over at larger radii as seen for IC~3639, NGC~788, and NGC~1667. The HST images for these sources are substantially deeper than some of the previously analyzed targets, indicating the turnover is likely real as the AGN radiation diminishes. The mass profiles are bumpy due to the clumpy nature of the NLR emission, as seen in Figure \ref{fig:sample}. The mass outflow rates similarly tend to rise with radius, but exhibit even more stochastic variations due to the radial changes in the outflow velocities, which do not vary smoothly as in systems that are modeled using a biconical outflow geometry with a unique peak velocity (originally classified as ``outflow'' targets by \citealp{Fischer2013}). These results highlight the critical role of spatially-resolved observations and photoionization models for studying outflows, as these radial variations would be missed by global observations using a single integrated spectrum.

In the three slightly higher luminosity AGN (NGC~788, NGC~3393, and NGC~7682) we observe peak outflow rates in the range 0.4 - 13 $M_{\odot}$ yr$^{-1}$, which are similar to those in \cite{Revalski2021}. In contrast, the peak outflow rates for two of the three slightly lower luminosity AGN (IC~3639 and UGC~1395) and are $\lesssim$~0.2 $M_{\odot}$ yr$^{-1}$. NGC~1667 stands out as a low-luminosity AGN with a peak mass outflow rate of $\sim$1.3 $M_{\odot}$ yr$^{-1}$. In the case of NGC~3393, the STIS slit is very close to the minor axis, resulting in large uncertainties in the velocity deprojection values. This is a limitation of the model, as material is likely flowing radially outward with some component up and off of the disk, which would result in smaller deprojections as more of the velocity vector is along our line of sight. In either case, the peak outflow rate is a modest $\sim$0.5~$M_{\odot}$ yr$^{-1}$. In most cases the velocity deprojection factors are $\sim$1.5~$-$~2.2, resulting in similar upper and lower limits on the mass outflow rates.

\begin{figure*}
\vspace{-1em}
\centering
\includegraphics[width=0.49\textwidth, trim={2em 0em 1em 0em}, clip]{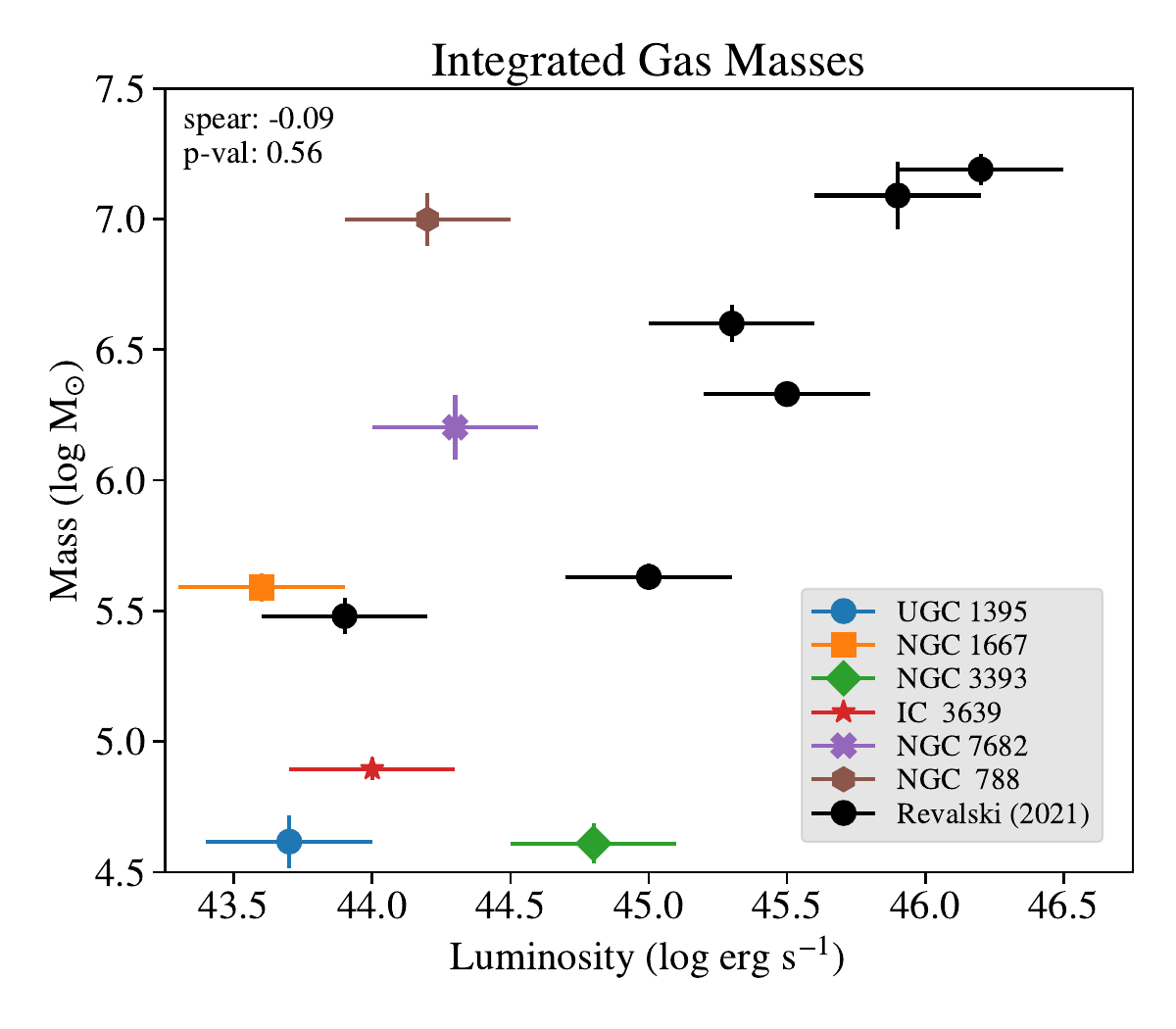}
\includegraphics[width=0.49\textwidth, trim={2em 0em 1em 0em}, clip]{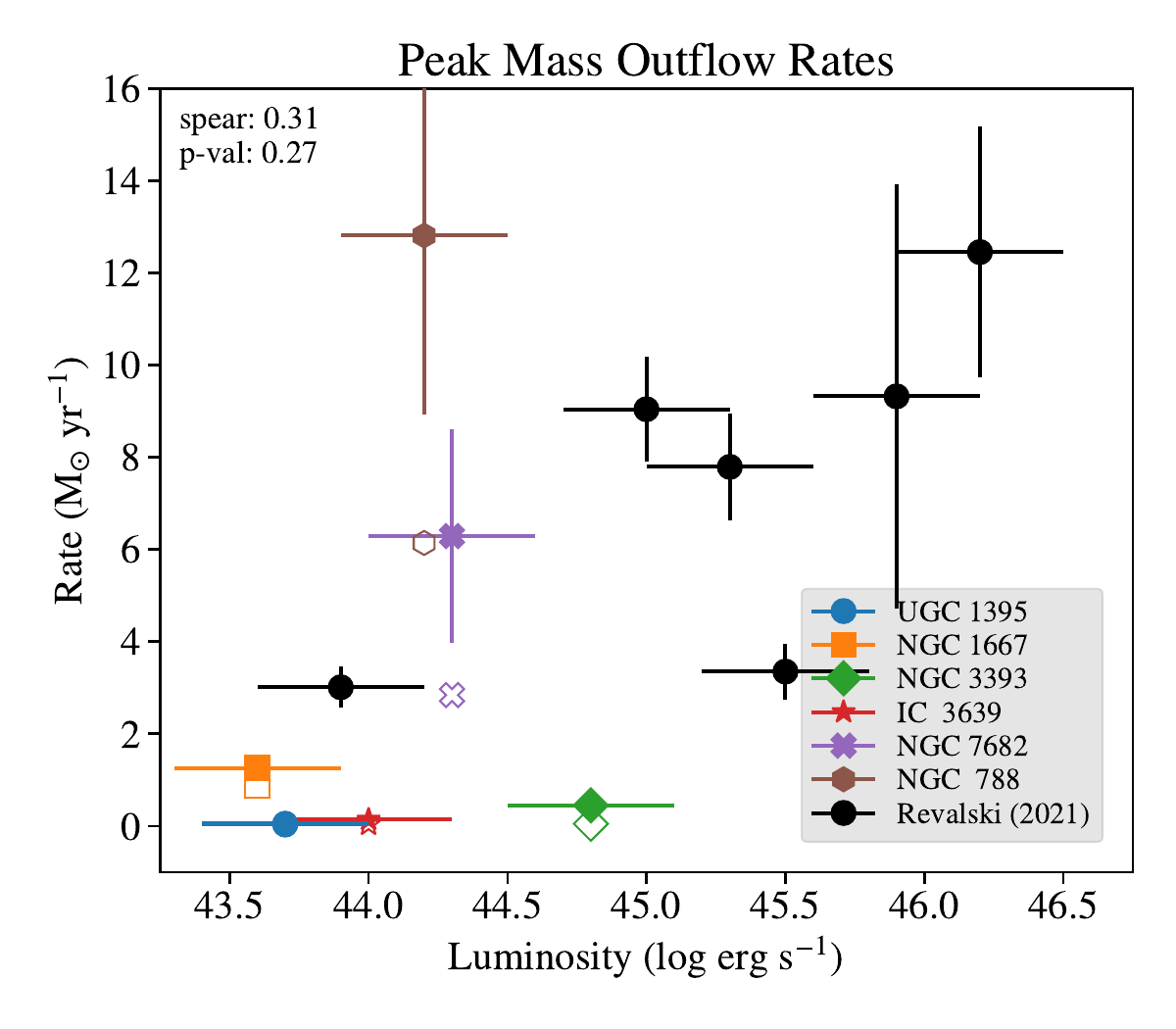}\\\vspace{-1.3em}
\includegraphics[width=0.49\textwidth, trim={2em 0em 1em 0em}, clip]{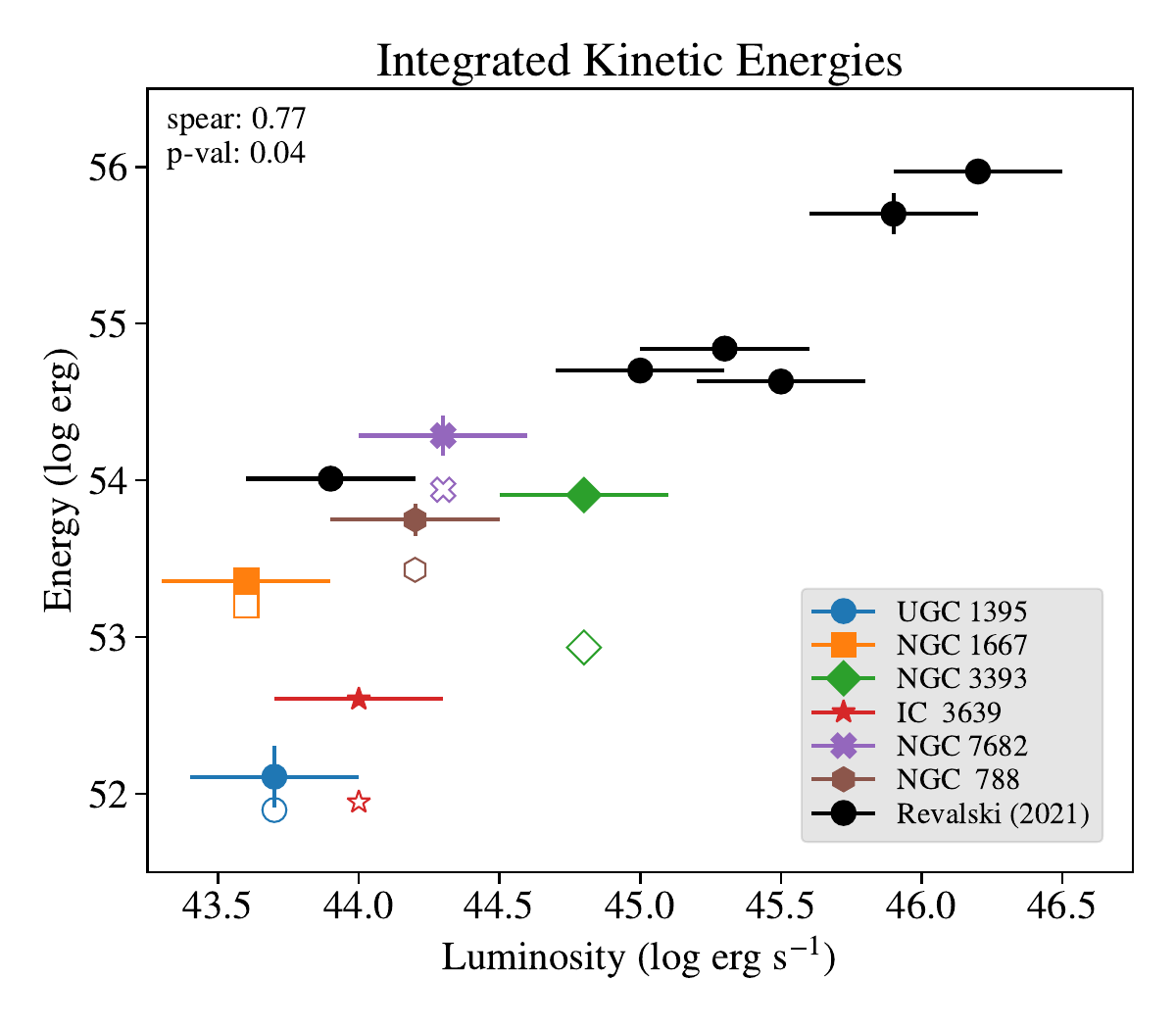}
\includegraphics[width=0.49\textwidth, trim={2em 0em 1em 0em}, clip]{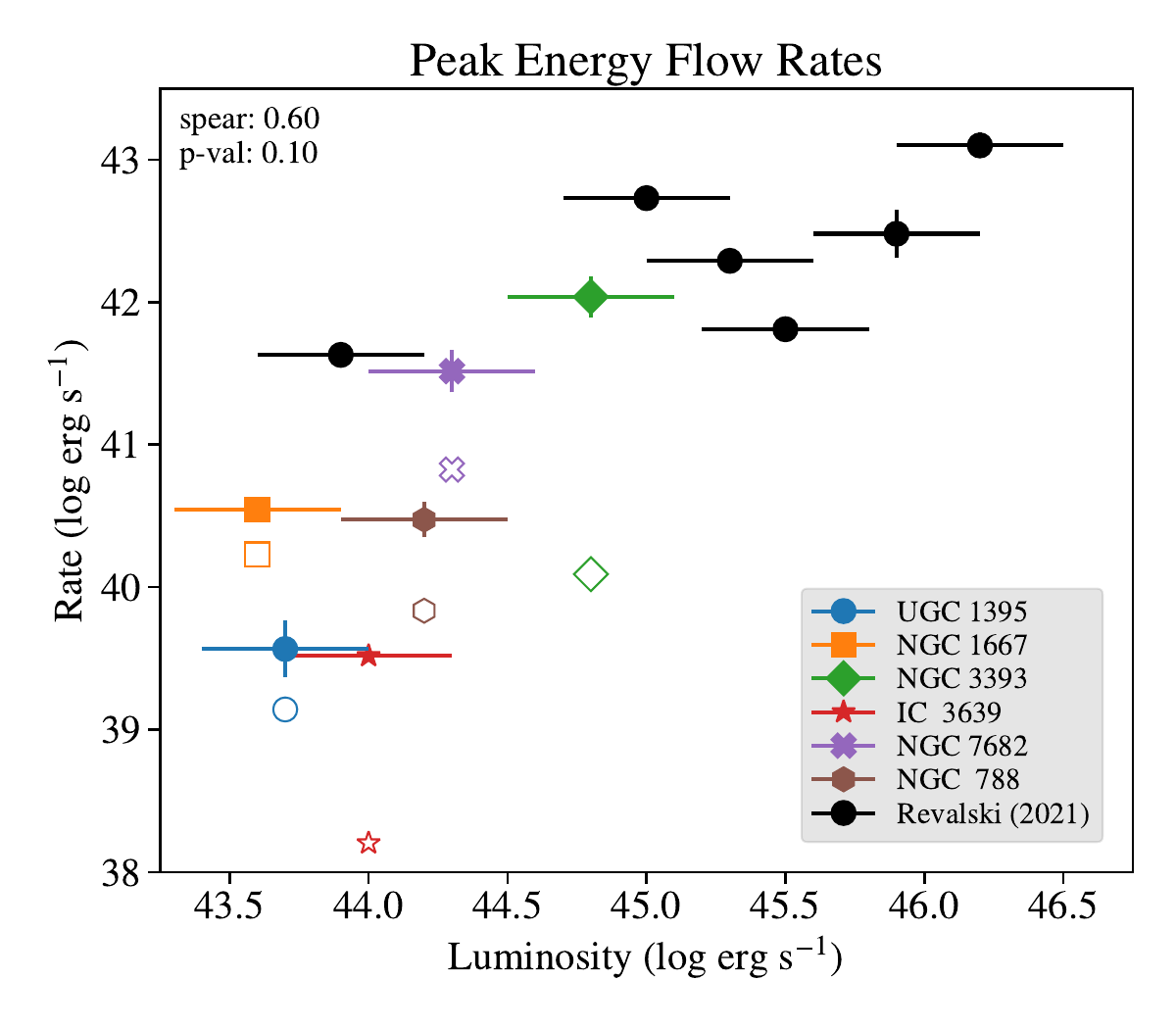}\\\vspace{-1.3em}
\includegraphics[width=0.49\textwidth, trim={2em 0em 1em 0em}, clip]{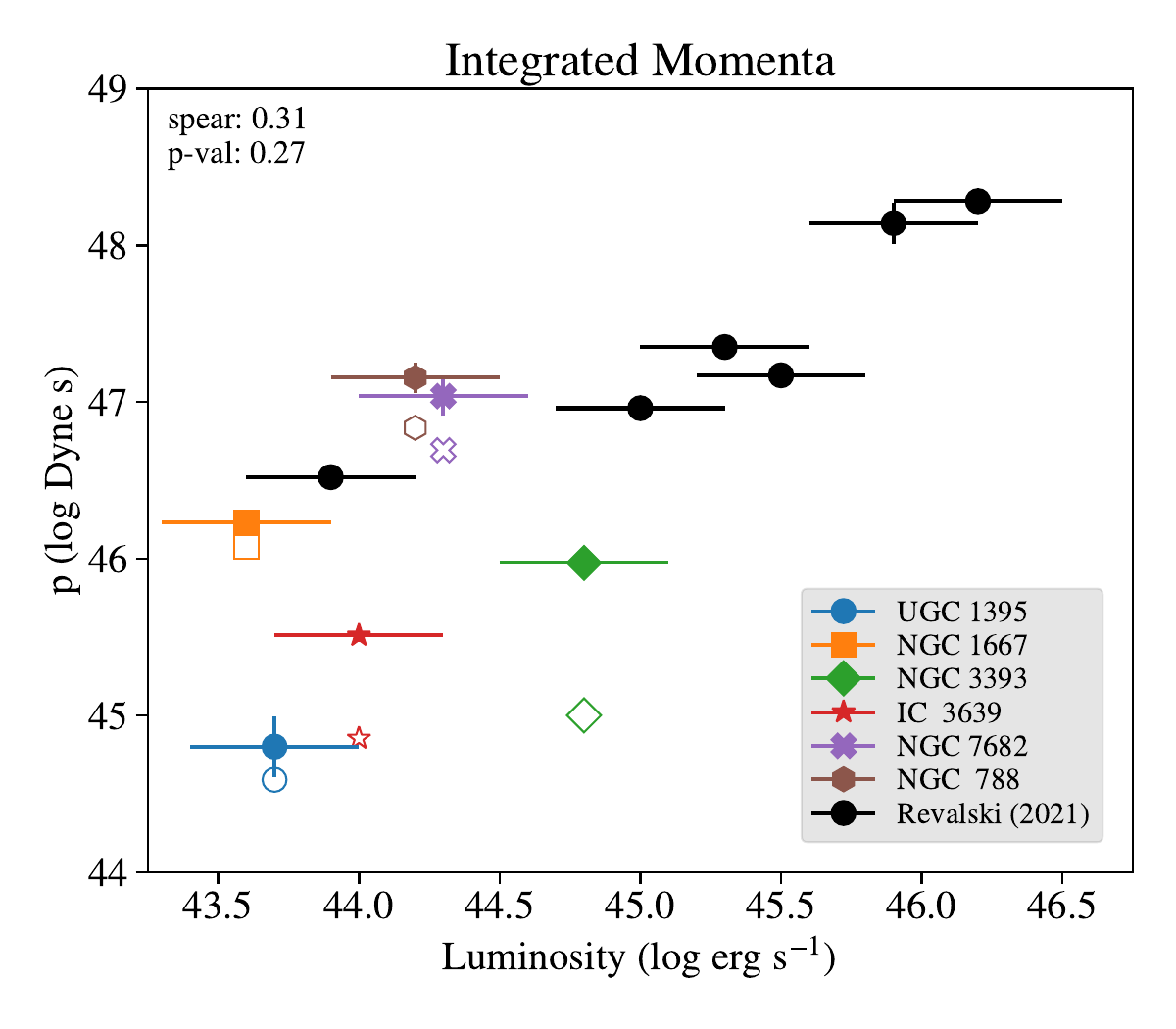}
\includegraphics[width=0.49\textwidth, trim={2em 0em 1em 0em}, clip]{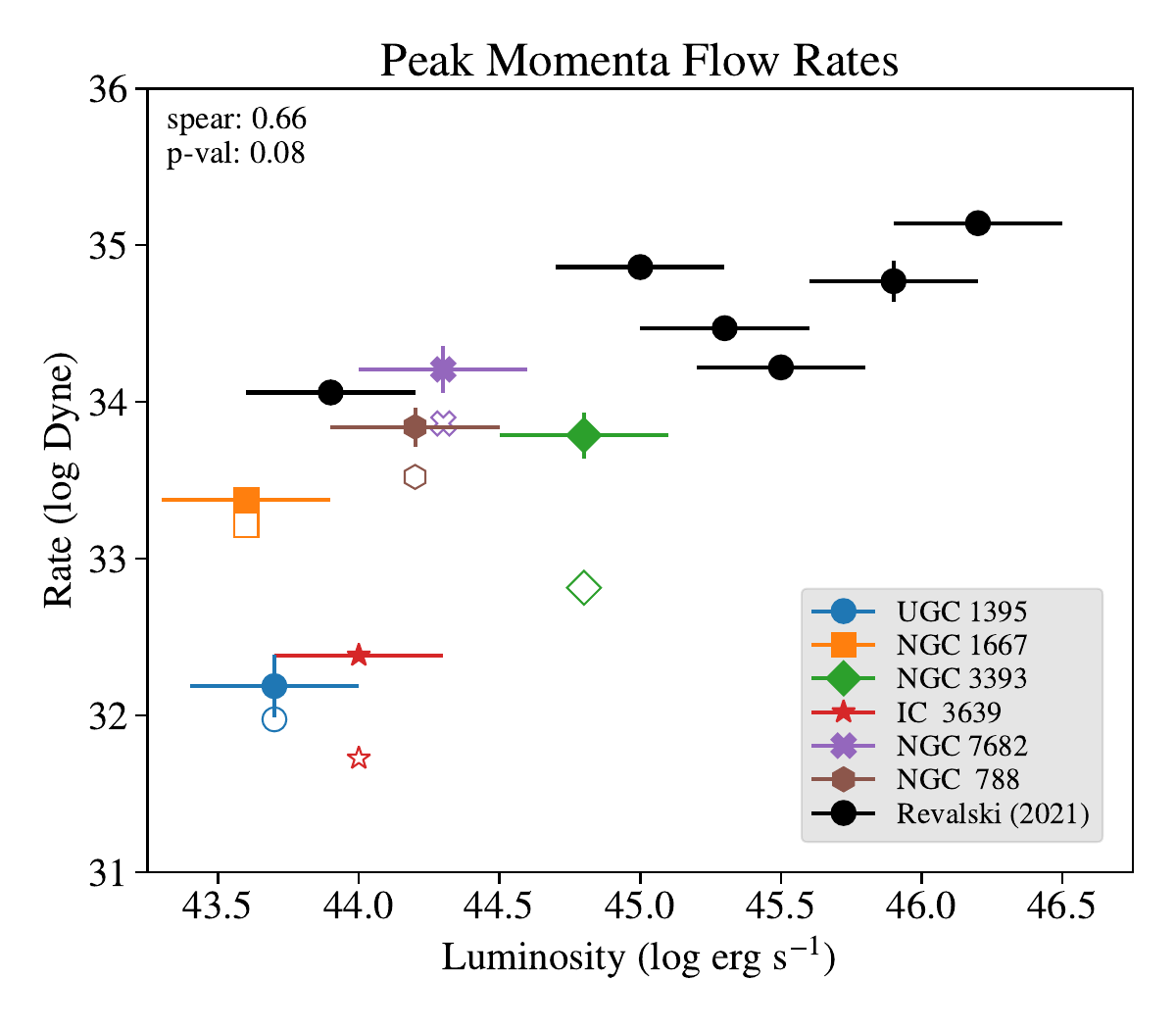}\vspace{-0.25em}\\
\vspace{-1.5em}
\caption{The integrated gas masses (upper left), kinetic energies (mid-left), and momenta (lower left) of the outflows, together with their peak outflow rates (right panels), as functions of the AGN bolometric luminosity for the six galaxies in this sample, as well as the six galaxies from \cite{Revalski2021}. The filled and open symbols represent the value for each energetic quantity calculated using the deprojected and observed velocities, respectively. There are clear correlations, with the Spearman's rank coefficient and p-value listed in the upper left of each panel for the deprojected quantities. The strongest correlations are found for total kinetic energy, peak energy flow rate, and peak momenta flow rate (p~$\leq$~0.1).}
\label{fig:results2}
\end{figure*}

We present the total ionized gas masses and kinetic energies integrated over all radial distances, as well as their peak flow rates, as a function of bolometric luminosity in Figure~\ref{fig:results2}. These results include only the outflowing gas, and exclude gas that is in rotation based on the kinematic analysis. As detailed in \cite{Revalski2018a}, the conserved quantities of mass, energy, and momentum may be spatially summed to obtain totals for the outflows, while the corresponding rates cannot be summed because they are sampling flow over discrete boundaries. For example, having observations with twice the current spatial resolution would result in twice as many outflow rate measurements, but the total mass participating in the outflow, and the amount of mass passing a given radius at any one time, are invariant to the spatial sampling.

The integrated masses are similar to those in \cite{Revalski2021}, with moderately higher gas masses in NGC~788 and NGC~7682 for their luminosities. The majority of AGN in the current study are lower in luminosity than the six targets from \cite{Revalski2021}, and we see correspondingly smaller kinetic energies and peak outflow rates. While there is scatter in these quantities, there are clear correlations with bolometric luminosity, particularly for the integrated outflow kinetic energy and peak energy outflow rates. We calculated the Spearman's rank correlation coefficient and corresponding p-values using the \href{https://docs.scipy.org/doc/scipy/reference/generated/scipy.stats.spearmanr.html}{stats.spearmanr} function within \href{https://www.scipy.org}{SciPy} (v1.12) and report these values on each panel of Figure~\ref{fig:results2}. We calculated the potential for feedback by comparing with theoretical models \citep{Hopkins2005}, and find that $\dot E$/\lbol~is $<$ 0.02\% in all cases, below the 0.5\%~$-$~5\% threshold required by simulations. However, we note that these models are most applicable to galaxies that are still assembling their components at higher redshifts.

The substantial radial variations in the gas masses are driven by intrinsic variations in the gas reservoirs at each radial distance, combined with the fraction of gas that is observed in the [O~III] ionization state. The energetics are more complex, as they also depend on the outflow velocities, which exhibit substantial changes with radius. These changes may be driven by a variety of complex factors including where the gas was launched from within the galaxy disk, the ionization state of the gas from launch to its current position, as well as the presence of obstacles such as spiral arms or slower moving outflows, possibly present in multiple gas phases. Considering these factors, the total integrated energetics, as well as their peak values, provide different but complimentary measures of the total and localized outflow energetics.

Critically, we observe that the mass outflow rate profiles are relatively constant or rising with radius, and may continue at larger radii where the [O~III] surface brightness falls below the sensitivity of our spectra and imaging. Thus, the total energetics presented in Figure~\ref{fig:results2} represent each value integrated to a given radius, and are lower limits on the total outflow energetics at all radii. These radial variations and lack of turnovers in the outflow rates highlight the necessity of using deep, spatially-resolved observations to obtain accurate measures of outflowing gas masses, velocities, and energetics.

\section{Discussion}
\label{sec:discussion}

A key result from this series of papers is a better understanding of how outflow densities change as a function of distance from the central SMBH. We have found high densities of $n_\mathrm{H} \approx 10^6$~cm$^{-3}$ in the nuclear regions of these AGN where most of the emission is produced, decreasing to $n_\mathrm{H} \approx 10^1$~cm$^{-3}$ at large radii. We have shown that assuming a constant density across the NLR outflows is not a valid assumption, and proven that it can lead to systematic uncertainties reaching several orders of magnitude. This has also been found by other detailed studies of gas densities \citep{Baron2019b, Davies2020Ric}, especially those that utilize high spatial resolution observations \citep{Dall'AgnoldeOliveira2021} and photoionization models \citep{Baron2019b, TrindadeFalcao2021}. Thus, it is absolutely essential to account for the radial variation in gas density when characterizing NLR gas and mass outflows.

Significant effort is being devoted to understanding AGN-driven outflows, including measuring their energetics and impacts on their host galaxy systems (e.g. \citealp{Fischer2018, Baron2019a, Baron2019b, ForsterSchreiber2019, Mingozzi2019, Davies2020Rebecca, Wylezalek2020, Avery2021, Fluetsch2021, Lamperti2021, Luo2021, Negus2021, Ruschel-Dutra2021, Speranza2021, Vayner2021, Bianchin2022, Deconto-Machado2022, Kakkad2022}). We find that the outflow energetics are comparable to or less than benchmarks for effective feedback from theoretical models, but in these local galaxies the evacuation of gas and thermal injection of energy may dominate the long term evolution of each system. Specifically, these AGN appear to be in ``maintenance mode" \citep{Gatto2024}, where the gas is kinematically disturbed from rotation, and heating and turbulence in the gas generated by the AGN radiation field may dominate any effects on star-formation in the galaxy on long time scales.

Critically, we note that measuring the extents of the ionized gas and outflows are limited by the sensitivity of detecting the H$\beta$ emission line. In our full sample, AGN-ionized [O~III] emission, as well as H$\alpha$ tracing turbulent kinematics, extend to larger radii. While [O~III]/H$\alpha$ can also provide a constraint on ionization of the gas, H$\beta$ or another recombination line is required for an accurate reddening correction. We recommend that future studies of AGN outflows utilize deeper spectroscopy, with S/N requirements to detect extended H$\beta$.

\begin{deluxetable*}{lccccccccc}
\tabletypesize{\small}
\setlength{\tabcolsep}{0.03in} 
\tablecaption{Integrated and Peak Outflow Properties}
\tablehead{
\colhead{Catalog} & \colhead{$\log$(\lbol)} & \colhead{Total $M$} &\colhead{$\dot M_{max}$} & \colhead{Total $E$} & \colhead{$\dot E_{max}$} & \colhead{Total $p$} & \colhead{$\dot p_{max}$} & \colhead{$\dot E_{max}$/\lbol} & \colhead{$\dot pc$/\lbol \vspace{-0.5em}}\\
\colhead{Name} & \colhead{(erg s$^{-1}$)} & \colhead{($\log M_{\odot}$)} &\colhead{($M_{\odot}$ yr$^{-1}$)} & \colhead{($\log$ erg)} & \colhead{($\log$ erg s$^{-1}$)} & \colhead{($\log$ dyne s)} & \colhead{($\log$ dyne)} & \colhead{(\%)} & \colhead{(\%) \vspace{-0.5em}}\\
\colhead{(1)} & \colhead{(2)} & \colhead{(3)} &\colhead{(4)} & \colhead{(5)} & \colhead{(6)} & \colhead{(7)} & \colhead{(8)} & \colhead{(9)} & \colhead{(10)}
}
\startdata
IC 3639  & 44.0 $\pm$ 0.3 & 4.90 $\pm$ 0.04 & 0.14 $\pm$ 0.01 & 52.60 $\pm$ 0.04 & 39.52 $\pm$ 0.04 & 45.51 $\pm$ 0.04 & 32.38 $\pm$ 0.04 & 0.003 $\pm$ 0.001 & 7 $\pm$ 1 \\
NGC 788  & 44.2 $\pm$ 0.3  & 7.00 $\pm$ 0.10 & 12.81 $\pm$ 3.88 & 53.75 $\pm$ 0.10 & 40.47 $\pm$ 0.12 & 47.15 $\pm$ 0.10 & 33.84 $\pm$ 0.12 & 0.019 $\pm$ 0.006 & 131 $\pm$ 40 \\
NGC 1667 & 43.6 $\pm$ 0.3  & 5.59 $\pm$ 0.06 & 1.26 $\pm$ 0.21 & 53.36 $\pm$ 0.05 & 40.54 $\pm$ 0.07 & 46.23 $\pm$ 0.05 & 33.37 $\pm$ 0.07 & 0.088 $\pm$ 0.015 & 178 $\pm$ 30 \\
NGC 3393 & 44.8 $\pm$ 0.3  & 4.61  $\pm$0.08 & 0.45 $\pm$ 0.08 & 53.91 $\pm$ 0.11 & 42.04 $\pm$ 0.15 & 45.97 $\pm$ 0.09 & 33.79 $\pm$ 0.15 & 0.173 $\pm$ 0.063 & 29 $\pm$ 11 \\
NGC 7682 & 44.3 $\pm$ 0.3  & 6.20 $\pm$ 0.12 & 6.29 $\pm$ 2.32 & 54.28 $\pm$ 0.13 & 41.52 $\pm$ 0.15 & 47.04 $\pm$ 0.13 & 34.21 $\pm$ 0.15 & 0.164 $\pm$ 0.061 & 242 $\pm$ 89 \\
UGC 1395 & 43.7 $\pm$ 0.3  & 4.62 $\pm$ 0.10 & 0.05 $\pm$ 0.03 & 52.11 $\pm$ 0.20 & 39.57 $\pm$ 0.20 & 44.80 $\pm$ 0.19 & 32.19 $\pm$ 0.20 & 0.007 $\pm$ 0.004 & 9 $\pm$ 5 \\ \hline
NGC 4151	 & 	43.9	$\pm$	0.3	 & 	5.48	$\pm$	0.07	 & 	3.01	$\pm$	0.45	 & 	54.01	$\pm$	0.08	 & 	41.63	$\pm$	0.09	 & 	46.52	$\pm$	0.08	 & 	34.06	$\pm$	0.08	 & 	0.54	$\pm$	0.11	 & 	437	$\pm$	85	 \\ 
NGC 1068	 & 	45.0	$\pm$	0.3	 & 	5.63	$\pm$	0.05	 & 	9.04	$\pm$	1.13	 & 	54.70	$\pm$	0.05	 & 	42.73	$\pm$	0.04	 & 	46.96	$\pm$	0.05	 & 	34.86	$\pm$	0.05	 & 	0.54	$\pm$	0.05	 & 	214	$\pm$	22	 \\ 
Mrk 3	 & 	45.3	$\pm$	0.3	 & 	6.60	$\pm$	0.07	 & 	7.79	$\pm$	1.15	 & 	54.84	$\pm$	0.06	 & 	42.29	$\pm$	0.07	 & 	47.35	$\pm$	0.06	 & 	34.47	$\pm$	0.06	 & 	0.10	$\pm$	0.02	 & 	44	$\pm$	6	 \\ 
Mrk 573	 & 	45.5	$\pm$	0.3	 & 	6.33	$\pm$	0.05	 & 	3.35	$\pm$	0.60	 & 	54.63	$\pm$	0.05	 & 	41.81	$\pm$	0.07	 & 	47.17	$\pm$	0.05	 & 	34.22	$\pm$	0.05	 & 	0.02	$\pm$	0.01	 & 	15	$\pm$	1	 \\ 
Mrk 78	 & 	45.9	$\pm$	0.3	 & 	7.09	$\pm$	0.13	 & 	9.32	$\pm$	4.60	 & 	55.70	$\pm$	0.13	 & 	42.48	$\pm$	0.17	 & 	48.14	$\pm$	0.13	 & 	34.77	$\pm$	0.13	 & 	0.04	$\pm$	0.02	 & 	22	$\pm$	6	 \\ 
Mrk 34	 & 	46.2	$\pm$	0.3	 & 	7.19	$\pm$	0.06	 & 	12.45	$\pm$	2.72	 & 	55.97	$\pm$	0.06	 & 	43.10	$\pm$	0.09	 & 	48.28	$\pm$	0.06	 & 	35.15	$\pm$	0.06	 & 	0.08	$\pm$	0.02	 & 	26	$\pm$	3
\enddata
\tablecomments{The tabulated results from Figure~\ref{fig:results2}. The columns are (1) target name, (2) bolometric luminosity, (3) total mass of the ionized gas, (4) peak mass outflow rate, (5) total kinetic energy, (6) peak kinetic energy rate, (7) total momentum, (8) peak momentum rate, (9) peak kinetic energy rate divided by the bolometric luminosity (percentage), and (10) the peak momentum rate divided by the photon momentum (percentage). Results for the six targets from \cite{Revalski2021} are shown below the horizontal line for completeness.}
\label{tab:results}
\end{deluxetable*}

In addition, the galaxies examined in this study highlight the important role of outflow geometry in defining the gas outflow velocities, with different models yielding moderately different results. The observed values always provide a lower limit on the outflow velocities and energetics, while a geometric model deprojects all or some portion of the velocity that is not directly along our line of sight. In the broadest picture, the results of \cite{Fischer2013} suggest that $\sim$1/3 of nearby Seyfert galaxy AGN show NLR outflows, defined by a relatively smooth rise and decline in velocity along the edges of a bicone.
Our study of six additional targets, falling primarily in Fischer et al.'s categories of complex or ambigious, reveal that they all also show NLR outflows \citep{Polack2024}.
Additional HST STIS spectroscopy is required for the remaining targets in their sample with archival observations that poorly sample the NLR emission in order to further constrain these statistics.
It may be that all AGN above a certain luminosity and Eddington ratio threshold show NLR outflows when observed with sufficient angular and spectral resolution, signal-to-noise, and spatial coverage, but this hypothesis remains to be tested by future studies of larger samples.

\newpage
\section{Conclusions}
\label{sec:conclusions}

We calculated spatially resolved gas masses and outflow rates for the ionized NLR outflows in six nearby active galaxies that have gas kinematics consistent with outflow along a galactic disk within the ionization bicone. Our main conclusions are the following:

\begin{enumerate}
\item The full sample of 12 galaxies from this series of studies spans 10$^3$ in bolometric luminosity. We find that the outflows contain total ionized gas masses of $M \approx 10^{4.6} - 10^{7.2}$ $M_{\odot}$, reach maximum mass outflow rates of $\dot M_{out} \approx 0.1 - 13$ $M_{\odot}$ yr$^{-1}$, and encompass total kinetic energies of $E \approx 10^{52} - 10^{56}$ erg. These properties all positively correlate with AGN luminosity.

\item While the radial outflow rate profiles are more stochastic than in biconical systems, they still exhibit a general rise, peak, and decline with radius. The differences between observed and deprojected velocities and outflow rates are often factors of a few, but can be higher, stressing the importance of modeling the geometry of each outflowing system when comparing galaxies.

\item The outflow energetics are less than benchmarks for effective feedback from theoretical models. However, these local galaxies exhibit maintenance mode feedback, where the thermal injection of energy from the AGN into the host galaxy interstellar medium may dominate long term effects on star-formation on large scales.

\item Our ability to measure ionized gas and outflow extents is limited by the sensitivity of detecting the H$\beta$ emission line. AGN-ionized [O~III] and H$\alpha$ emission tracing turbulent kinematics often extend to larger radii where there may be significant mass. We recommend that future studies of AGN outflows utilize deeper spectroscopy, with S/N requirements to detect extended H$\beta$.

\item The results for these six galaxies, together with the six galaxies that we characterized in \cite{Revalski2021}, represent the full sample of low-redshift Seyfert galaxies from \cite{Fischer2013, Fischer2014} with sufficient HST STIS spectroscopy and [O~III] imaging required for our spatially-resolved modeling technique. A larger sample is needed to fully explore the dependence of spatially-resolved outflow properties on host galaxy and AGN properties, including a wider range in luminosity.
\end{enumerate}

\acknowledgments

The authors thank the anonymous referee for helpful comments that improved the clarity of this study. Based on observations with the NASA/ESA Hubble Space Telescope obtained from the MAST Data Archive at the Space Telescope Science Institute, which is operated by the Association of Universities for Research in Astronomy, Incorporated, under NASA contract NAS5-26555. Support for program numbers 16246 was provided through a grant from the STScI under NASA contract NAS5-26555. These observations are associated with program numbers \href{https://archive.stsci.edu/proposal_search.php?mission=hst&id=16246}{16246}. The WFC3 observations can be accessed using the DOI: \dataset[10.17909/np2s-gd34]{https://doi.org/10.17909/np2s-gd34}. We also provide our custom calibrated mosaics for use by the community as High Level Science Products available from the Mikulski Archive for Space Telescopes (MAST) using the DOI: \dataset[10.17909/7ccz-mc93]{https://doi.org/10.17909/7ccz-mc93}, and directly online at \url{https://archive.stsci.edu/hlsp/nlr-agn/}.

This paper used the photoionization code Cloudy, which can be obtained from \url{http://www.nublado.org} and the Atomic Line List available at \url{http://www.pa.uky.edu/~peter/atomic/}. This research has made use of the NASA/IPAC Extragalactic Database (NED), which is operated by the Jet Propulsion Laboratory, California Institute of Technology, under contract with the National Aeronautics and Space Administration. This research has made use of NASA 's Astrophysics Data System.

\facilities{HST(STIS, WFC3)}

\software{Astropy \citep{AstropyCollaboration2013, AstropyCollaboration2018, AstropyCollaboration2022}, IPython \citep{Perez2007}, Jupyter \citep{Kluyver2016}, Matplotlib \citep{Hunter2007, Caswell2021}, NumPy \citep{Harris2020}, Python (\citealp{VanRossum2009}, \url{https://www.python.org}), Scipy \citep{Virtanen2020a, Virtanen2020b}, MultiNest \citep{Feroz2019}, 
Cloudy \citep{Ferland2013, Chatzikos2023}}\\


\bibliography{references}{}

\begin{thebibliography}{}
\expandafter\ifx\csname natexlab\endcsname\relax\def\natexlab#1{#1}\fi
\providecommand{\url}[1]{\href{#1}{#1}}
\providecommand{\dodoi}[1]{doi:~\href{http://doi.org/#1}{\nolinkurl{#1}}}
\providecommand{\doeprint}[1]{\href{http://ascl.net/#1}{\nolinkurl{http://ascl.net/#1}}}
\providecommand{\doarXiv}[1]{\href{https://arxiv.org/abs/#1}{\nolinkurl{https://arxiv.org/abs/#1}}}

\bibitem[{{Astropy Collaboration} {et~al.}(2013){Astropy Collaboration}, {Robitaille}, {Tollerud}, {Greenfield}, {Droettboom}, {Bray}, {Aldcroft}, {Davis}, {Ginsburg}, {Price-Whelan}, {Kerzendorf}, {Conley}, {Crighton}, {Barbary}, {Muna}, {Ferguson}, {Grollier}, {Parikh}, {Nair}, {Unther}, {Deil}, {Woillez}, {Conseil}, {Kramer}, {Turner}, {Singer}, {Fox}, {Weaver}, {Zabalza}, {Edwards}, {Azalee Bostroem}, {Burke}, {Casey}, {Crawford}, {Dencheva}, {Ely}, {Jenness}, {Labrie}, {Lim}, {Pierfederici}, {Pontzen}, {Ptak}, {Refsdal}, {Servillat}, \& {Streicher}}]{AstropyCollaboration2013}
{Astropy Collaboration}, {Robitaille}, T.~P., {Tollerud}, E.~J., {et~al.} 2013, \aap, 558, A33, \dodoi{10.1051/0004-6361/201322068}

\bibitem[{{Astropy Collaboration} {et~al.}(2018){Astropy Collaboration}, {Price-Whelan}, {Sip{\H{o}}cz}, {G{\"u}nther}, {Lim}, {Crawford}, {Conseil}, {Shupe}, {Craig}, {Dencheva}, {Ginsburg}, {VanderPlas}, {Bradley}, {P{\'e}rez-Su{\'a}rez}, {de Val-Borro}, {Aldcroft}, {Cruz}, {Robitaille}, {Tollerud}, {Ardelean}, {Babej}, {Bach}, {Bachetti}, {Bakanov}, {Bamford}, {Barentsen}, {Barmby}, {Baumbach}, {Berry}, {Biscani}, {Boquien}, {Bostroem}, {Bouma}, {Brammer}, {Bray}, {Breytenbach}, {Buddelmeijer}, {Burke}, {Calderone}, {Cano Rodr{\'\i}guez}, {Cara}, {Cardoso}, {Cheedella}, {Copin}, {Corrales}, {Crichton}, {D'Avella}, {Deil}, {Depagne}, {Dietrich}, {Donath}, {Droettboom}, {Earl}, {Erben}, {Fabbro}, {Ferreira}, {Finethy}, {Fox}, {Garrison}, {Gibbons}, {Goldstein}, {Gommers}, {Greco}, {Greenfield}, {Groener}, {Grollier}, {Hagen}, {Hirst}, {Homeier}, {Horton}, {Hosseinzadeh}, {Hu}, {Hunkeler}, {Ivezi{\'c}}, {Jain}, {Jenness}, {Kanarek}, {Kendrew}, {Kern}, {Kerzendorf}, {Khvalko}, {King}, {Kirkby}, {Kulkarni},
  {Kumar}, {Lee}, {Lenz}, {Littlefair}, {Ma}, {Macleod}, {Mastropietro}, {McCully}, {Montagnac}, {Morris}, {Mueller}, {Mumford}, {Muna}, {Murphy}, {Nelson}, {Nguyen}, {Ninan}, {N{\"o}the}, {Ogaz}, {Oh}, {Parejko}, {Parley}, {Pascual}, {Patil}, {Patil}, {Plunkett}, {Prochaska}, {Rastogi}, {Reddy Janga}, {Sabater}, {Sakurikar}, {Seifert}, {Sherbert}, {Sherwood-Taylor}, {Shih}, {Sick}, {Silbiger}, {Singanamalla}, {Singer}, {Sladen}, {Sooley}, {Sornarajah}, {Streicher}, {Teuben}, {Thomas}, {Tremblay}, {Turner}, {Terr{\'o}n}, {van Kerkwijk}, {de la Vega}, {Watkins}, {Weaver}, {Whitmore}, {Woillez}, {Zabalza}, \& {Astropy Contributors}}]{AstropyCollaboration2018}
{Astropy Collaboration}, {Price-Whelan}, A.~M., {Sip{\H{o}}cz}, B.~M., {et~al.} 2018, \aj, 156, 123, \dodoi{10.3847/1538-3881/aabc4f}

\bibitem[{{Astropy Collaboration} {et~al.}(2022){Astropy Collaboration}, {Price-Whelan}, {Lim}, {Earl}, {Starkman}, {Bradley}, {Shupe}, {Patil}, {Corrales}, {Brasseur}, {N{\"o}the}, {Donath}, {Tollerud}, {Morris}, {Ginsburg}, {Vaher}, {Weaver}, {Tocknell}, {Jamieson}, {van Kerkwijk}, {Robitaille}, {Merry}, {Bachetti}, {G{\"u}nther}, {Aldcroft}, {Alvarado-Montes}, {Archibald}, {B{\'o}di}, {Bapat}, {Barentsen}, {Baz{\'a}n}, {Biswas}, {Boquien}, {Burke}, {Cara}, {Cara}, {Conroy}, {Conseil}, {Craig}, {Cross}, {Cruz}, {D'Eugenio}, {Dencheva}, {Devillepoix}, {Dietrich}, {Eigenbrot}, {Erben}, {Ferreira}, {Foreman-Mackey}, {Fox}, {Freij}, {Garg}, {Geda}, {Glattly}, {Gondhalekar}, {Gordon}, {Grant}, {Greenfield}, {Groener}, {Guest}, {Gurovich}, {Handberg}, {Hart}, {Hatfield-Dodds}, {Homeier}, {Hosseinzadeh}, {Jenness}, {Jones}, {Joseph}, {Kalmbach}, {Karamehmetoglu}, {Ka{\l}uszy{\'n}ski}, {Kelley}, {Kern}, {Kerzendorf}, {Koch}, {Kulumani}, {Lee}, {Ly}, {Ma}, {MacBride}, {Maljaars}, {Muna}, {Murphy}, {Norman},
  {O'Steen}, {Oman}, {Pacifici}, {Pascual}, {Pascual-Granado}, {Patil}, {Perren}, {Pickering}, {Rastogi}, {Roulston}, {Ryan}, {Rykoff}, {Sabater}, {Sakurikar}, {Salgado}, {Sanghi}, {Saunders}, {Savchenko}, {Schwardt}, {Seifert-Eckert}, {Shih}, {Jain}, {Shukla}, {Sick}, {Simpson}, {Singanamalla}, {Singer}, {Singhal}, {Sinha}, {Sip{\H{o}}cz}, {Spitler}, {Stansby}, {Streicher}, {{\v{S}}umak}, {Swinbank}, {Taranu}, {Tewary}, {Tremblay}, {de Val-Borro}, {Van Kooten}, {Vasovi{\'c}}, {Verma}, {de Miranda Cardoso}, {Williams}, {Wilson}, {Winkel}, {Wood-Vasey}, {Xue}, {Yoachim}, {Zhang}, {Zonca}, \& {Astropy Project Contributors}}]{AstropyCollaboration2022}
{Astropy Collaboration}, {Price-Whelan}, A.~M., {Lim}, P.~L., {et~al.} 2022, \apj, 935, 167, \dodoi{10.3847/1538-4357/ac7c74}

\bibitem[{{Avery} {et~al.}(2021){Avery}, {Wuyts}, {F{\"o}rster Schreiber}, {Villforth}, {Bertemes}, {Chang}, {Hamer}, {Toshikawa}, \& {Zhang}}]{Avery2021}
{Avery}, C.~R., {Wuyts}, S., {F{\"o}rster Schreiber}, N.~M., {et~al.} 2021, \mnras, 503, 5134, \dodoi{10.1093/mnras/stab780}

\bibitem[{{Baldwin} {et~al.}(1981){Baldwin}, {Phillips}, \& {Terlevich}}]{Baldwin1981}
{Baldwin}, J.~A., {Phillips}, M.~M., \& {Terlevich}, R. 1981, \pasp, 93, 5, \dodoi{10.1086/130766}

\bibitem[{{Baron} \& {Netzer}(2019{\natexlab{a}})}]{Baron2019a}
{Baron}, D., \& {Netzer}, H. 2019{\natexlab{a}}, \mnras, 482, 3915, \dodoi{10.1093/mnras/sty2935}

\bibitem[{{Baron} \& {Netzer}(2019{\natexlab{b}})}]{Baron2019b}
---. 2019{\natexlab{b}}, \mnras, 486, 4290, \dodoi{10.1093/mnras/stz1070}

\bibitem[{{Bianchin} {et~al.}(2022){Bianchin}, {Riffel}, {Storchi-Bergmann}, {Riffel}, {Ruschel-Dutra}, {Harrison}, {Dahmer-Hahn}, {Mainieri}, {Sch{\"o}nell}, \& {Dametto}}]{Bianchin2022}
{Bianchin}, M., {Riffel}, R.~A., {Storchi-Bergmann}, T., {et~al.} 2022, \mnras, 510, 639, \dodoi{10.1093/mnras/stab3468}

\bibitem[{{Bischetti} {et~al.}(2017){Bischetti}, {Piconcelli}, {Vietri}, {Bongiorno}, {Fiore}, {Sani}, {Marconi}, {Duras}, {Zappacosta}, {Brusa}, {Comastri}, {Cresci}, {Feruglio}, {Giallongo}, {La Franca}, {Mainieri}, {Mannucci}, {Martocchia}, {Ricci}, {Schneider}, {Testa}, \& {Vignali}}]{Bischetti2017}
{Bischetti}, M., {Piconcelli}, E., {Vietri}, G., {et~al.} 2017, \aap, 598, A122, \dodoi{10.1051/0004-6361/201629301}

\bibitem[{{Caswell} {et~al.}(2021){Caswell}, {Droettboom}, {Lee}, {Sales de Andrade}, {Hoffmann}, {Hunter}, {Klymak}, {Firing}, {Stansby}, {Varoquaux}, {Hedegaard Nielsen}, {Root}, {May}, {Elson}, {Sepp{\"a}nen}, {Dale}, {Lee}, {McDougall}, {Straw}, {Hobson}, {hannah}, {Gohlke}, {Vincent}, {Yu}, {Ma}, {Silvester}, {Moad}, {Kniazev}, {Ernest}, \& {Ivanov}}]{Caswell2021}
{Caswell}, T.~A., {Droettboom}, M., {Lee}, A., {et~al.} 2021, {matplotlib/matplotlib: REL: v3.5.0}, v3.5.0, Zenodo,  Zenodo, \dodoi{10.5281/zenodo.5706396}

\bibitem[{{Chatzikos} {et~al.}(2023){Chatzikos}, {Bianchi}, {Camilloni}, {Chakraborty}, {Gunasekera}, {Guzm{\'a}n}, {Milby}, {Sarkar}, {Shaw}, {van Hoof}, \& {Ferland}}]{Chatzikos2023}
{Chatzikos}, M., {Bianchi}, S., {Camilloni}, F., {et~al.} 2023, \rmxaa, 59, 327, \dodoi{10.22201/ia.01851101p.2023.59.02.12}

\bibitem[{{Ciotti} \& {Ostriker}(2001)}]{Ciotti2001}
{Ciotti}, L., \& {Ostriker}, J.~P. 2001, \apj, 551, 131, \dodoi{10.1086/320053}

\bibitem[{{Collins} {et~al.}(2009){Collins}, {Kraemer}, {Crenshaw}, {Bruhweiler}, \& {Mel{\'e}ndez}}]{Collins2009}
{Collins}, N.~R., {Kraemer}, S.~B., {Crenshaw}, D.~M., {Bruhweiler}, F.~C., \& {Mel{\'e}ndez}, M. 2009, \apj, 694, 765, \dodoi{10.1088/0004-637X/694/2/765}

\bibitem[{{Cresci} \& {Maiolino}(2018)}]{Cresci2018}
{Cresci}, G., \& {Maiolino}, R. 2018, Nature Astronomy, 2, 179, \dodoi{10.1038/s41550-018-0404-5}

\bibitem[{{Dall'Agnol de Oliveira} {et~al.}(2021){Dall'Agnol de Oliveira}, {Storchi-Bergmann}, {Kraemer}, {Villar Mart{\'\i}n}, {Schnorr-M{\"u}ller}, {Schmitt}, {Ruschel-Dutra}, {Crenshaw}, \& {Fischer}}]{Dall'AgnoldeOliveira2021}
{Dall'Agnol de Oliveira}, B., {Storchi-Bergmann}, T., {Kraemer}, S.~B., {et~al.} 2021, \mnras, 504, 3890, \dodoi{10.1093/mnras/stab1067}

\bibitem[{{Davies} {et~al.}(2020{\natexlab{a}}){Davies}, {Baron}, {Shimizu}, {Netzer}, {Burtscher}, {de Zeeuw}, {Genzel}, {Hicks}, {Koss}, {Lin}, {Lutz}, {Maciejewski}, {M{\"u}ller-S{\'a}nchez}, {Orban de Xivry}, {Ricci}, {Riffel}, {Riffel}, {Rosario}, {Schartmann}, {Schnorr-M{\"u}ller}, {Shangguan}, {Sternberg}, {Sturm}, {Storchi-Bergmann}, {Tacconi}, \& {Veilleux}}]{Davies2020Ric}
{Davies}, R., {Baron}, D., {Shimizu}, T., {et~al.} 2020{\natexlab{a}}, \mnras, 498, 4150, \dodoi{10.1093/mnras/staa2413}

\bibitem[{{Davies} {et~al.}(2020{\natexlab{b}}){Davies}, {F{\"o}rster Schreiber}, {Lutz}, {Genzel}, {Belli}, {Shimizu}, {Contursi}, {Davies}, {Herrera-Camus}, {Lee}, {Naab}, {Price}, {Renzini}, {Schruba}, {Sternberg}, {Tacconi}, {{\"U}bler}, {Wisnioski}, \& {Wuyts}}]{Davies2020Rebecca}
{Davies}, R.~L., {F{\"o}rster Schreiber}, N.~M., {Lutz}, D., {et~al.} 2020{\natexlab{b}}, \apj, 894, 28, \dodoi{10.3847/1538-4357/ab86ad}

\bibitem[{{Deconto-Machado} {et~al.}(2022){Deconto-Machado}, {Riffel}, {Ilha}, {Rembold}, {Storchi-Bergmann}, {Riffel}, {Schimoia}, {Schneider}, {Bizyaev}, {Feng}, {Wylezalek}, {da Costa}, {do Nascimento}, \& {Maia}}]{Deconto-Machado2022}
{Deconto-Machado}, A., {Riffel}, R.~A., {Ilha}, G.~S., {et~al.} 2022, \aap, 659, A131, \dodoi{10.1051/0004-6361/202140613}

\bibitem[{{Ferland} {et~al.}(2013){Ferland}, {Porter}, {van Hoof}, {Williams}, {Abel}, {Lykins}, {Shaw}, {Henney}, \& {Stancil}}]{Ferland2013}
{Ferland}, G.~J., {Porter}, R.~L., {van Hoof}, P.~A.~M., {et~al.} 2013, \rmxaa, 49, 137.
\newblock \doarXiv{1302.4485}

\bibitem[{{Feroz} {et~al.}(2019){Feroz}, {Hobson}, {Cameron}, \& {Pettitt}}]{Feroz2019}
{Feroz}, F., {Hobson}, M.~P., {Cameron}, E., \& {Pettitt}, A.~N. 2019, The Open Journal of Astrophysics, 2, 10, \dodoi{10.21105/astro.1306.2144}

\bibitem[{{Fiore} {et~al.}(2017){Fiore}, {Feruglio}, {Shankar}, {Bischetti}, {Bongiorno}, {Brusa}, {Carniani}, {Cicone}, {Duras}, {Lamastra}, {Mainieri}, {Marconi}, {Menci}, {Maiolino}, {Piconcelli}, {Vietri}, \& {Zappacosta}}]{Fiore2017}
{Fiore}, F., {Feruglio}, C., {Shankar}, F., {et~al.} 2017, \aap, 601, A143, \dodoi{10.1051/0004-6361/201629478}

\bibitem[{{Fischer} {et~al.}(2013){Fischer}, {Crenshaw}, {Kraemer}, \& {Schmitt}}]{Fischer2013}
{Fischer}, T.~C., {Crenshaw}, D.~M., {Kraemer}, S.~B., \& {Schmitt}, H.~R. 2013, \apjs, 209, 1, \dodoi{10.1088/0067-0049/209/1/1}

\bibitem[{{Fischer} {et~al.}(2014){Fischer}, {Crenshaw}, {Kraemer}, {Schmitt}, \& {Turner}}]{Fischer2014}
{Fischer}, T.~C., {Crenshaw}, D.~M., {Kraemer}, S.~B., {Schmitt}, H.~R., \& {Turner}, T.~J. 2014, \apj, 785, 25, \dodoi{10.1088/0004-637X/785/1/25}

\bibitem[{{Fischer} {et~al.}(2017){Fischer}, {Machuca}, {Diniz}, {Crenshaw}, {Kraemer}, {Riffel}, {Schmitt}, {Baron}, {Storchi-Bergmann}, {Straughn}, {Revalski}, \& {Pope}}]{Fischer2017}
{Fischer}, T.~C., {Machuca}, C., {Diniz}, M.~R., {et~al.} 2017, \apj, 834, 30, \dodoi{10.3847/1538-4357/834/1/30}

\bibitem[{{Fischer} {et~al.}(2018){Fischer}, {Kraemer}, {Schmitt}, {Longo Micchi}, {Crenshaw}, {Revalski}, {Vestergaard}, {Elvis}, {Gaskell}, {Hamann}, {Ho}, {Hutchings}, {Mushotzky}, {Netzer}, {Storchi-Bergmann}, {Straughn}, {Turner}, \& {Ward}}]{Fischer2018}
{Fischer}, T.~C., {Kraemer}, S.~B., {Schmitt}, H.~R., {et~al.} 2018, \apj, 856, 102, \dodoi{10.3847/1538-4357/aab03e}

\bibitem[{{Florez} {et~al.}(2021){Florez}, {Jogee}, {Guo}, {Cora}, {Weinberger}, {Dav{\'e}}, {Hernquist}, {Vogelsberger}, {Ciardullo}, {Finkelstein}, {Gronwall}, {Kawinwanichakij}, {Leung}, {LaMassa}, {Papovich}, {Stevans}, \& {Wold}}]{Florez2021}
{Florez}, J., {Jogee}, S., {Guo}, Y., {et~al.} 2021, \mnras, 508, 762, \dodoi{10.1093/mnras/stab2593}

\bibitem[{{Fluetsch} {et~al.}(2021){Fluetsch}, {Maiolino}, {Carniani}, {Arribas}, {Belfiore}, {Bellocchi}, {Cazzoli}, {Cicone}, {Cresci}, {Fabian}, {Gallagher}, {Ishibashi}, {Mannucci}, {Marconi}, {Perna}, {Sturm}, \& {Venturi}}]{Fluetsch2021}
{Fluetsch}, A., {Maiolino}, R., {Carniani}, S., {et~al.} 2021, \mnras, 505, 5753, \dodoi{10.1093/mnras/stab1666}

\bibitem[{{F{\"o}rster Schreiber} {et~al.}(2019){F{\"o}rster Schreiber}, {{\"U}bler}, {Davies}, {Genzel}, {Wisnioski}, {Belli}, {Shimizu}, {Lutz}, {Fossati}, {Herrera-Camus}, {Mendel}, {Tacconi}, {Wilman}, {Beifiori}, {Brammer}, {Burkert}, {Carollo}, {Davies}, {Eisenhauer}, {Fabricius}, {Lilly}, {Momcheva}, {Naab}, {Nelson}, {Price}, {Renzini}, {Saglia}, {Sternberg}, {van Dokkum}, \& {Wuyts}}]{ForsterSchreiber2019}
{F{\"o}rster Schreiber}, N.~M., {{\"U}bler}, H., {Davies}, R.~L., {et~al.} 2019, \apj, 875, 21, \dodoi{10.3847/1538-4357/ab0ca2}

\bibitem[{{Gatto} {et~al.}(2024){Gatto}, {Storchi-Bergmann}, {Riffel}, {Riffel}, {Rembold}, {Schimoia}, {Mallmann}, \& {Ilha}}]{Gatto2024}
{Gatto}, L., {Storchi-Bergmann}, T., {Riffel}, R.~A., {et~al.} 2024, \mnras, 530, 3059, \dodoi{10.1093/mnras/stae989}

\bibitem[{{Gnilka} {et~al.}(2020){Gnilka}, {Crenshaw}, {Fischer}, {Revalski}, {Meena}, {Martinez}, {Polack}, {Machuca}, {Dashtamirova}, {Kraemer}, {Schmitt}, {Riffel}, \& {Storchi-Bergmann}}]{Gnilka2020}
{Gnilka}, C.~L., {Crenshaw}, D.~M., {Fischer}, T.~C., {et~al.} 2020, \apj, 893, 80, \dodoi{10.3847/1538-4357/ab8000}

\bibitem[{{Harris} {et~al.}(2020){Harris}, {Millman}, {van der Walt}, {Gommers}, {Virtanen}, {Cournapeau}, {Wieser}, {Taylor}, {Berg}, {Smith}, {Kern}, {Picus}, {Hoyer}, {van Kerkwijk}, {Brett}, {Haldane}, {del R{\'\i}o}, {Wiebe}, {Peterson}, {G{\'e}rard-Marchant}, {Sheppard}, {Reddy}, {Weckesser}, {Abbasi}, {Gohlke}, \& {Oliphant}}]{Harris2020}
{Harris}, C.~R., {Millman}, K.~J., {van der Walt}, S.~J., {et~al.} 2020, \nat, 585, 357, \dodoi{10.1038/s41586-020-2649-2}

\bibitem[{{Harrison} {et~al.}(2018){Harrison}, {Costa}, {Tadhunter}, {Fl{\"u}tsch}, {Kakkad}, {Perna}, \& {Vietri}}]{Harrison2018}
{Harrison}, C.~M., {Costa}, T., {Tadhunter}, C.~N., {et~al.} 2018, Nature Astronomy, 2, 198, \dodoi{10.1038/s41550-018-0403-6}

\bibitem[{{Heckman} \& {Best}(2014)}]{Heckman2014}
{Heckman}, T.~M., \& {Best}, P.~N. 2014, \araa, 52, 589, \dodoi{10.1146/annurev-astro-081913-035722}

\bibitem[{{Hopkins} {et~al.}(2005){Hopkins}, {Hernquist}, {Cox}, {Di Matteo}, {Martini}, {Robertson}, \& {Springel}}]{Hopkins2005}
{Hopkins}, P.~F., {Hernquist}, L., {Cox}, T.~J., {et~al.} 2005, \apj, 630, 705, \dodoi{10.1086/432438}

\bibitem[{{Hunter}(2007)}]{Hunter2007}
{Hunter}, J.~D. 2007, Computing in Science and Engineering, 9, 90, \dodoi{10.1109/MCSE.2007.55}

\bibitem[{{Kakkad} {et~al.}(2022){Kakkad}, {Sani}, {Rojas}, {Mallmann}, {Veilleux}, {Bauer}, {Ricci}, {Mushotzky}, {Koss}, {Ricci}, {Treister}, {Privon}, {Nguyen}, {B{\"a}r}, {Harrison}, {Oh}, {Powell}, {Riffel}, {Stern}, {Trakhtenbrot}, \& {Urry}}]{Kakkad2022}
{Kakkad}, D., {Sani}, E., {Rojas}, A.~F., {et~al.} 2022, \mnras, 511, 2105, \dodoi{10.1093/mnras/stac103}

\bibitem[{{Karouzos} {et~al.}(2016){Karouzos}, {Woo}, \& {Bae}}]{Karouzos2016}
{Karouzos}, M., {Woo}, J.-H., \& {Bae}, H.-J. 2016, \apj, 833, 171, \dodoi{10.3847/1538-4357/833/2/171}

\bibitem[{{Kauffmann} {et~al.}(2003){Kauffmann}, {Heckman}, {Tremonti}, {Brinchmann}, {Charlot}, {White}, {Ridgway}, {Brinkmann}, {Fukugita}, {Hall}, {Ivezi{\'c}}, {Richards}, \& {Schneider}}]{Kauffmann2003}
{Kauffmann}, G., {Heckman}, T.~M., {Tremonti}, C., {et~al.} 2003, \mnras, 346, 1055, \dodoi{10.1111/j.1365-2966.2003.07154.x}

\bibitem[{{Kewley} {et~al.}(2001){Kewley}, {Dopita}, {Sutherland}, {Heisler}, \& {Trevena}}]{Kewley2001}
{Kewley}, L.~J., {Dopita}, M.~A., {Sutherland}, R.~S., {Heisler}, C.~A., \& {Trevena}, J. 2001, \apj, 556, 121, \dodoi{10.1086/321545}

\bibitem[{{Kewley} {et~al.}(2006){Kewley}, {Groves}, {Kauffmann}, \& {Heckman}}]{Kewley2006}
{Kewley}, L.~J., {Groves}, B., {Kauffmann}, G., \& {Heckman}, T. 2006, \mnras, 372, 961, \dodoi{10.1111/j.1365-2966.2006.10859.x}

\bibitem[{{Kluyver} {et~al.}(2016){Kluyver}, {Ragan-Kelley}, {P{\'e}rez}, {Granger}, {Bussonnier}, {Frederic}, {Kelley}, {Hamrick}, {Grout}, {Corlay}, {Ivanov}, {Avila}, {Abdalla}, {Willing}, \& {Jupyter Development Team}}]{Kluyver2016}
{Kluyver}, T., {Ragan-Kelley}, B., {P{\'e}rez}, F., {et~al.} 2016, in IOS Press (IOS Press), 87--90, \dodoi{10.3233/978-1-61499-649-1-87}

\bibitem[{{Kormendy} \& {Ho}(2013)}]{Kormendy2013}
{Kormendy}, J., \& {Ho}, L.~C. 2013, \araa, 51, 511, \dodoi{10.1146/annurev-astro-082708-101811}

\bibitem[{{Laha} {et~al.}(2021){Laha}, {Reynolds}, {Reeves}, {Kriss}, {Guainazzi}, {Smith}, {Veilleux}, \& {Proga}}]{Laha2021}
{Laha}, S., {Reynolds}, C.~S., {Reeves}, J., {et~al.} 2021, Nature Astronomy, 5, 13, \dodoi{10.1038/s41550-020-01255-2}

\bibitem[{{Lamperti} {et~al.}(2021){Lamperti}, {Harrison}, {Mainieri}, {Kakkad}, {Perna}, {Circosta}, {Scholtz}, {Carniani}, {Cicone}, {Alexander}, {Bischetti}, {Calistro Rivera}, {Chen}, {Cresci}, {Feruglio}, {Fiore}, {Mannucci}, {Marconi}, {Mart{\'\i}nez-Ram{\'\i}rez}, {Netzer}, {Piconcelli}, {Puglisi}, {Rosario}, {Schramm}, {Vietri}, {Vignali}, \& {Zappacosta}}]{Lamperti2021}
{Lamperti}, I., {Harrison}, C.~M., {Mainieri}, V., {et~al.} 2021, \aap, 654, A90, \dodoi{10.1051/0004-6361/202141363}

\bibitem[{{Luo} {et~al.}(2021){Luo}, {Woo}, {Karouzos}, {Bae}, {Shin}, {McConnell}, {Shih}, {Kim}, \& {Park}}]{Luo2021}
{Luo}, R., {Woo}, J.-H., {Karouzos}, M., {et~al.} 2021, \apj, 908, 221, \dodoi{10.3847/1538-4357/abd5ac}

\bibitem[{{Mingozzi} {et~al.}(2019){Mingozzi}, {Cresci}, {Venturi}, {Marconi}, {Mannucci}, {Perna}, {Belfiore}, {Carniani}, {Balmaverde}, {Brusa}, {Cicone}, {Feruglio}, {Gallazzi}, {Mainieri}, {Maiolino}, {Nagao}, {Nardini}, {Sani}, {Tozzi}, \& {Zibetti}}]{Mingozzi2019}
{Mingozzi}, M., {Cresci}, G., {Venturi}, G., {et~al.} 2019, \aap, 622, A146, \dodoi{10.1051/0004-6361/201834372}

\bibitem[{{Negus} {et~al.}(2021){Negus}, {Comerford}, {M{\"u}ller S{\'a}nchez}, {Barrera-Ballesteros}, {Drory}, {Rembold}, \& {Riffel}}]{Negus2021}
{Negus}, J., {Comerford}, J.~M., {M{\"u}ller S{\'a}nchez}, F., {et~al.} 2021, \apj, 920, 62, \dodoi{10.3847/1538-4357/ac1343}

\bibitem[{{Osterbrock} \& {Ferland}(2006)}]{Osterbrock2006}
{Osterbrock}, D.~E., \& {Ferland}, G.~J. 2006, {Astrophysics of gaseous nebulae and active galactic nuclei} (University Science Books)

\bibitem[{{Perez} \& {Granger}(2007)}]{Perez2007}
{Perez}, F., \& {Granger}, B.~E. 2007, Computing in Science and Engineering, 9, 21, \dodoi{10.1109/MCSE.2007.53}

\bibitem[{{Perna} {et~al.}(2017){Perna}, {Lanzuisi}, {Brusa}, {Cresci}, \& {Mignoli}}]{Perna2017}
{Perna}, M., {Lanzuisi}, G., {Brusa}, M., {Cresci}, G., \& {Mignoli}, M. 2017, \aap, 606, A96, \dodoi{10.1051/0004-6361/201730819}

\bibitem[{{Polack} {et~al.}(2024){Polack}, {Revalski}, {Crenshaw}, {Fischer}, {Schmitt}, {Kraemer}, {Meena}, \& {Rafelski}}]{Polack2024}
{Polack}, G.~E., {Revalski}, M., {Crenshaw}, D.~M., {et~al.} 2024, \apj, 975, 129, \dodoi{10.3847/1538-4357/ad71c3}

\bibitem[{{Revalski} {et~al.}(2020){Revalski}, {Crenshaw}, {Fischer}, {Kraemer}, {Meena}, {Polack}, {Rafelski}, \& {Schmitt}}]{Revalski2020}
{Revalski}, M., {Crenshaw}, D.~M., {Fischer}, T., {et~al.} 2020, {Are Narrow Line Region Outflows an Effective Mode of AGN Feedback?}, HST Proposal. Cycle 28, ID. \#16246

\bibitem[{{Revalski} {et~al.}(2018){Revalski}, {Crenshaw}, {Kraemer}, {Fischer}, {Schmitt}, \& {Machuca}}]{Revalski2018a}
{Revalski}, M., {Crenshaw}, D.~M., {Kraemer}, S.~B., {et~al.} 2018, \apj, 856, 46, \dodoi{10.3847/1538-4357/aab107}

\bibitem[{{Revalski} {et~al.}(2021){Revalski}, {Meena}, {Martinez}, {Polack}, {Crenshaw}, {Kraemer}, {Collins}, {Fischer}, {Schmitt}, {Schmidt}, {Maksym}, \& {Rafelski}}]{Revalski2021}
{Revalski}, M., {Meena}, B., {Martinez}, F., {et~al.} 2021, \apj, 910, 139, \dodoi{10.3847/1538-4357/abdcad}

\bibitem[{{Revalski} {et~al.}(2022){Revalski}, {Crenshaw}, {Rafelski}, {Kraemer}, {Polack}, {Falc{\~a}o}, {Fischer}, {Meena}, {Martinez}, {Schmitt}, {Collins}, \& {Falcone}}]{Revalski2022}
{Revalski}, M., {Crenshaw}, D.~M., {Rafelski}, M., {et~al.} 2022, \apj, 930, 14, \dodoi{10.3847/1538-4357/ac5f3d}

\bibitem[{{Riess} {et~al.}(2022){Riess}, {Yuan}, {Macri}, {Scolnic}, {Brout}, {Casertano}, {Jones}, {Murakami}, {Anand}, {Breuval}, {Brink}, {Filippenko}, {Hoffmann}, {Jha}, {D'arcy Kenworthy}, {Mackenty}, {Stahl}, \& {Zheng}}]{Riess2022}
{Riess}, A.~G., {Yuan}, W., {Macri}, L.~M., {et~al.} 2022, \apjl, 934, L7, \dodoi{10.3847/2041-8213/ac5c5b}

\bibitem[{{Ruschel-Dutra} {et~al.}(2021){Ruschel-Dutra}, {Storchi-Bergmann}, {Schnorr-M{\"u}ller}, {Riffel}, {Dall'Agnol de Oliveira}, {Lena}, {Robinson}, {Nagar}, \& {Elvis}}]{Ruschel-Dutra2021}
{Ruschel-Dutra}, D., {Storchi-Bergmann}, T., {Schnorr-M{\"u}ller}, A., {et~al.} 2021, \mnras, 507, 74, \dodoi{10.1093/mnras/stab2058}

\bibitem[{{Seab} \& {Shull}(1983)}]{Seab1983}
{Seab}, C.~G., \& {Shull}, J.~M. 1983, \apj, 275, 652, \dodoi{10.1086/161563}

\bibitem[{{Snow} \& {Witt}(1996)}]{Snow1996}
{Snow}, T.~P., \& {Witt}, A.~N. 1996, \apjl, 468, L65, \dodoi{10.1086/310225}

\bibitem[{{Speranza} {et~al.}(2021){Speranza}, {Balmaverde}, {Capetti}, {Massaro}, {Tremblay}, {Marconi}, {Venturi}, {Chiaberge}, {Baldi}, {Baum}, {Grandi}, {Meyer}, {O'Dea}, {Sparks}, {Terrazas}, \& {Torresi}}]{Speranza2021}
{Speranza}, G., {Balmaverde}, B., {Capetti}, A., {et~al.} 2021, \aap, 653, A150, \dodoi{10.1051/0004-6361/202140686}

\bibitem[{{Storchi-Bergmann} \& {Schnorr-M{\"u}ller}(2019)}]{Storchi-Bergmann2019}
{Storchi-Bergmann}, T., \& {Schnorr-M{\"u}ller}, A. 2019, Nature Astronomy, 3, 48, \dodoi{10.1038/s41550-018-0611-0}

\bibitem[{{Trindade Falc{\~a}o} {et~al.}(2021){Trindade Falc{\~a}o}, {Kraemer}, {Fischer}, {Crenshaw}, {Revalski}, {Schmitt}, {Vestergaard}, {Elvis}, {Gaskell}, {Hamann}, {Ho}, {Hutchings}, {Mushotzky}, {Netzer}, {Storchi-Bergmann}, {Turner}, \& {Ward}}]{TrindadeFalcao2021}
{Trindade Falc{\~a}o}, A., {Kraemer}, S.~B., {Fischer}, T.~C., {et~al.} 2021, \mnras, 500, 1491, \dodoi{10.1093/mnras/staa3239}

\bibitem[{Van~Rossum \& Drake(2009)}]{VanRossum2009}
Van~Rossum, G., \& Drake, F.~L. 2009, Python 3 Reference Manual (Scotts Valley, CA: CreateSpace)

\bibitem[{{Vayner} {et~al.}(2021){Vayner}, {Zakamska}, {Wright}, {Armus}, {Murray}, \& {Walth}}]{Vayner2021}
{Vayner}, A., {Zakamska}, N., {Wright}, S.~A., {et~al.} 2021, \apj, 923, 59, \dodoi{10.3847/1538-4357/ac2b9e}

\bibitem[{{Veilleux} {et~al.}(2020){Veilleux}, {Maiolino}, {Bolatto}, \& {Aalto}}]{Veilleux2020}
{Veilleux}, S., {Maiolino}, R., {Bolatto}, A.~D., \& {Aalto}, S. 2020, \aapr, 28, 2, \dodoi{10.1007/s00159-019-0121-9}

\bibitem[{{Veilleux} \& {Osterbrock}(1987)}]{Veilleux1987}
{Veilleux}, S., \& {Osterbrock}, D.~E. 1987, \apjs, 63, 295, \dodoi{10.1086/191166}

\bibitem[{{Virtanen} {et~al.}(2020{\natexlab{a}}){Virtanen}, {Gommers}, {Oliphant}, {Haberland}, {Reddy}, {Cournapeau}, {Burovski}, {Peterson}, {Weckesser}, {Bright}, {van der Walt}, {Brett}, {Wilson}, {Millman}, {Mayorov}, {Nelson}, {Jones}, {Kern}, {Larson}, {Carey}, {Polat}, {Feng}, {Moore}, {VanderPlas}, {Laxalde}, {Perktold}, {Cimrman}, {Henriksen}, {Quintero}, {Harris}, {Archibald}, {Ribeiro}, {Pedregosa}, {van Mulbregt}, \& {SciPy 1. 0 Contributors}}]{Virtanen2020a}
{Virtanen}, P., {Gommers}, R., {Oliphant}, T.~E., {et~al.} 2020{\natexlab{a}}, Nature Methods, 17, 261, \dodoi{10.1038/s41592-019-0686-2}

\bibitem[{{Virtanen} {et~al.}(2020{\natexlab{b}}){Virtanen}, {Gommers}, {Burovski}, {Oliphant}, {Weckesser}, {Cournapeau}, {Alexbrc}, {Peterson}, {Wilson}, {Reddy}, {Mayorov}, {Endolith}, {Haberland}, {Nelson}, {Van Der Walt}, {Laxalde}, {Brett}, {Polat}, {Larson}, {Millman}, {Lars}, {Van Mulbregt}, {Eric-Jones}, {Carey}, {Moore}, {Kern}, {Leslie}, {Perktold}, {Striega}, \& {Feng}}]{Virtanen2020b}
{Virtanen}, P., {Gommers}, R., {Burovski}, E., {et~al.} 2020{\natexlab{b}}, {scipy/scipy: SciPy 1.5.2}, v1.5.2, Zenodo,  Zenodo, \dodoi{10.5281/zenodo.3958354}

\bibitem[{{Wylezalek} {et~al.}(2020){Wylezalek}, {Flores}, {Zakamska}, {Greene}, \& {Riffel}}]{Wylezalek2020}
{Wylezalek}, D., {Flores}, A.~M., {Zakamska}, N.~L., {Greene}, J.~E., \& {Riffel}, R.~A. 2020, \mnras, 492, 4680, \dodoi{10.1093/mnras/staa062}

\end{thebibliography}
\bibliographystyle{aasjournal}


\end{document}